\documentclass[aps,pra,twocolumn,export]{revtex4-1} 
\usepackage[colorlinks=true,linkcolor=blue,citecolor=blue,urlcolor=blue]{hyperref}

\usepackage[utf8]{inputenc}
\usepackage[german, english]{babel}

\usepackage{dcolumn}  
\usepackage{bm}        
\usepackage{bbold}
\usepackage{subfigure}
\usepackage{amsthm}
\usepackage{mathtools}

\usepackage[export]{adjustbox}
\usepackage{titlesec}
\usepackage{amsfonts}
\usepackage{amssymb}
\usepackage{braket}

\usepackage[usenames,dvipsnames]{xcolor}
 \definecolor{david}{rgb}{0.7,0,0.9}
 	
\definecolor{henrik}{rgb}{0.1,0.6,0.4}

\newcommand{\mbb}{\mathbb}
\newcommand{\mcal}{\mathcal}
\newcommand{\mbf}{\mathbf}

\begin{document}

\title{Detecting subsystem symmetry-protected topological order via entanglement entropy}
\author{David T. Stephen}
\author{Henrik Dreyer}
\author{Mohsin Iqbal}
\author{Norbert Schuch}
\affiliation{Max-Planck-Institut f{\"u}r Quantenoptik, Hans-Kopfermann-Stra{\ss}e 1, 85748 Garching, Germany}
\affiliation{Munich Center for Quantum Science and Technology, Schellingstra{\ss}e 4, 80799 M{\"u}nchen, Germany}
\date{\today}

\begin{abstract}
Subsystem symmetry-protected topological (SSPT) order is a type of quantum order that is protected by symmetries acting on lower-dimensional subsystems of the entire system. In this paper, we show how SSPT order can be characterized and detected by a constant correction to the entanglement area law, similar to the topological entanglement entropy. Focusing on the paradigmatic two-dimensional cluster phase as an example, we use tensor network methods to give an analytic argument that almost all states in the phase exhibit the same correction to the area law, such that this correction may be used to reliably detect the SSPT order of the cluster phase. Based on this idea, we formulate a numerical method that uses tensor networks to extract this correction from ground state wave functions. We use this method to study the fate of the SSPT order of the cluster state under various external fields and interactions, and find that the correction persists unless a phase transition is crossed, or the subsystem symmetry is explicitly broken. Surprisingly, these results uncover that the SSPT order of the cluster state persists beyond the cluster phase, thanks to a new type of subsystem time-reversal symmetry. Finally, we discuss the correction to the area law found in 3D cluster states on different lattices, indicating rich behaviour for general subsystem symmetries.

\end{abstract}

\maketitle

The modern perspective of quantum phases of matter is based on global patterns of entanglement \cite{Chen2010,Wen2017,Zeng2019}. Two quantum states are said to lie in distinct (symmetry-protected) topological phases when they cannot be connected by (symmetric) local unitary evolution. As such, topological phases of matter can be characterized and detected by entanglement-based quantities. A well-known example is the topological entanglement entropy (TEE): For states with non-trivial topological order, the scaling of entanglement entropy exhibits a correction to the area law \cite{Hamma2005,Levin2006,Kitaev2006,Eisert2010}. That is, for a subregion $A$ of a lattice, the entropy for ground states of gapped, local Hamiltonians takes the general form,
\begin{equation} \label{eq:entropy}
    S_A=a |\partial A| -\gamma +\dots
\end{equation}
for some constants $a,\gamma$, where $\partial A$ is the boundary of region $A$ and the dots indicate terms that decay exponentially with $|A|$. The correction $\gamma$ takes a uniform non-zero value within non-trivial topological phases as a consequence of the global entanglement patterns. It can therefore be used to detect and characterize topological order and topological phase transitions analytically, numerically, and potentially even experimentally \cite{Papanikolaou2007,Hamma2008,Nussinov2009,Isakov2011,Jiang2012,Flammia2009,Halasz2012,Abanin2012,Daley2012,Islam2015}.

Recently, however, it has been observed that $\gamma$ may deviate from the expected value due to the presence of symmetry-protected topological (SPT) order localized around $\partial A$ \cite{Cano2015,Zou2016,Santos2018,Devakul2018b,Williamson2019,Schmitz2019}. One setting in which this occurs is for states with subsystem SPT (SSPT) order \cite{Raussendorf2019,You2018,Devakul2018b,Devakul2019}. Such order is non-trivial only in the presence of subsystem symmetries, which are defined as symmetries that act on rigid lower-dimensional subsystems of the entire system. In cases where these symmetries act on  1D lines spanning a 2D lattice, one may find a non-zero value of $\gamma$ for regions $A$ whose boundaries lie parallel to these lines, despite the absence of topological order \cite{Zou2016,Devakul2018b,Williamson2019}.

On one hand, these corrections may be viewed as obstructions to reliably extracting the TEE from ground states, and previous research has been focused on developing methods to get around this \cite{Zou2016,Williamson2019}. On the other hand, they suggest the possibility that SSPT order may also be characterized by corrections to the area law. This is an attractive prospect, as SSPT order has recently garnered notable interest in both the contexts of condensed matter, thanks to its relation to fracton order \cite{Vijay2016,Williamson2016,Kubica2018,You2018a,Shirley2019,Song2019,Ma2018}, and quantum information, due to its use in measurement-based quantum computation \cite{Else2012a,Raussendorf2019,Stephen2019,Devakul2018a}. Thus, any tool that can be used to characterize and detect SSPT order would also have immediate impact in these areas.

Can $\gamma$ provide such a tool? Up to this point, SSPT order has been given as a sufficient condition for $\gamma\geq \log 2$ \cite{Devakul2018b}, but also an example of two states in the same SSPT phase with different non-zero values of $\gamma$ has been given \cite{Williamson2019}. This suggests that SSPT phases of matter are not associated with a particular value of $\gamma$, and therefore that the precise value of $\gamma$ is not useful for the characterization of SSPT order.
In this work, we demonstrate the contrary. We consider the 2D cluster phase, previously discussed in Refs.~\cite{Raussendorf2019,Else2012a,You2018,Devakul2018b}, and show that $\gamma$ takes the same value for all generic states in the phase, with deviations only occurring at fine-tuned points. 
This value of $\gamma$, which we refer to as the \textit{symmetry-protected entanglement entropy (SPEE)}, relates to the non-trivial symmetry fractionalization that occurs on the boundary of every state in the cluster phase.
Therefore, the SPEE may be used to characterize SSPT phases of matter in the same way that TEE characterizes topological phases of matter.

We use this result as the basis for a new numerical technique to detect SSPT order in ground states of gapped local Hamiltonians. Namely, we show how one can straightforwardly extract $\gamma$ from a projected entangled pair state (PEPS) representation of a ground state, and then apply this method to various Hamiltonians obtained by perturbing the cluster state Hamiltonian. In accordance with our analytical arguments, we find that $\gamma$ can reliably and unambiguously detect SSPT order and SSPT phase transitions. Moreover, since $\gamma$ makes no reference to any particular symmetry of the system, it detects any and all SSPT order within a ground state, unlike usual SPT order parameters which must be defined with respect to a specific symmetry \cite{Bartlett2010,Haegeman2012,Pollmann2012a,Marvian2017,You2018}. As a consequence of this, we discover a large region in which the SSPT order of the cluster state appears to persist despite the subsystem symmetries being explicitly broken. We make a preliminary analysis of this new phase of matter in terms of a new notion of subsystem time-reversal symmetry of the cluster state. 

Together, our analytical and numerical results show that the SPEE is an effective tool to detect and characterize the SSPT order of the 2D cluster phase. Going beyond this, we also study 3D cluster states with different types of subsystem symmetries and calculate $\gamma$ in each case, observing distinct behaviours depending on the structure of the symmetries. We believe that our uniformity arguments for the 2D cluster phase will hold equally well for these 3D phases, and also other types of SSPT order. In an outlook, we discuss the implications of our results for measurement-based quantum computation, detection of fracton order, and the possible experimental observation of SSPT order.

This paper is organized as follows. In Sec.~\ref{sec:cluster}, we define the cluster state and cluster phase, and show their relation to corrections to the area law. In Sec.~\ref{sec:anal}, we present an analytical argument that $\gamma$ is the same for almost every state in the cluster phase. Then, in Sec.~\ref{sec:num}, we formulate our numerical method and use it to examine the SSPT order of the cluster state with various Hamiltonian fields and interactions added. In Sec.~\ref{sec:3d}, we calculate the area law corrections for cluster states defined on various 3D lattices, and finally in Sec.~\ref{sec:con}, we present our conclusions and future directions of work.

\section{Cluster phase and corrections to the area law} \label{sec:cluster}

We will begin by reviewing the definition of the cluster state and the cluster phase. We will then calculate the entanglement entropy for a continuous one-parameter family of states in the cluster phase, and find that all states in this family have the same correction to the area law, except for a singular point where the correction is larger. We trace back this larger correction to extra symmetries of the state which do not generically hold in the cluster phase, thereby establishing the premise that all generic states in the cluster phase have the same correction to the area law.

\begin{figure}
    \centering
    \includegraphics[width=0.9\linewidth]{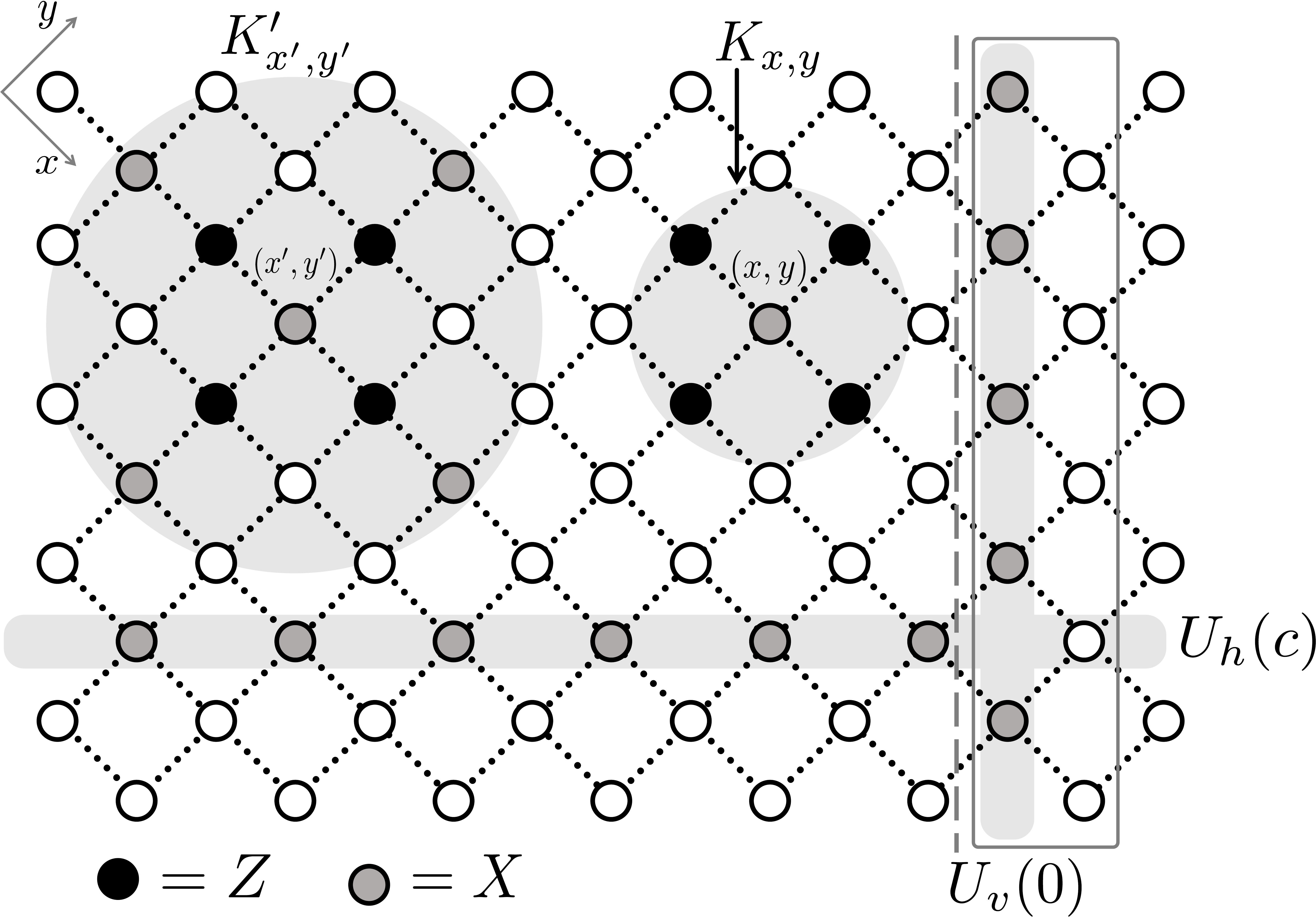}
    \caption{A section of the rotated square lattice considered in this paper. The lattice lives on an infinite cylinder, extending infinitely to the left and right with periodic boundary conditions in the vertical direction. Here, the circumference is $N=5$. The dashed line indicates the boundary between the $A$ (right) and $B$ (left) subsystems. The solid rectangle denotes the two columns of sites that make up one block in the quasi-1D system considered in Sec.~\ref{sec:anal}. The subsystem symmetries $U_{h,v}(c)$ defining the cluster phase are pictured, as are the stabilizers $K_{x,y}$ and $K'_{x,y}$ that define the states $|C\rangle$ and $|C(\pi)\rangle$, respectively.}
    \label{fig:symms}
\end{figure}

Throughout this paper, we consider the geometry of a $45^\circ$ rotated 2D square lattice on an infinitely long cylinder with circumference $N$ (although similar results hold also for a finite cylinder or torus with dimensions much larger than the correlation length). The cylinder is bipartitioned into two semi-infinite subsystems, denoted $A$ and $B$. The 2D cluster state $|C\rangle$ \cite{Raussendorf2001a,Raussendorf2003} can be defined on this lattice by placing a qubit in the $|+\rangle =\frac{1}{\sqrt{2}}(|0\rangle+|1\rangle)$ state on each lattice site, and applying the unitary $CZ=\mathrm{diag}(1,1,1,-1)$ between each pair of neighbouring qubits. It may equivalently be defined as the unique solution to the constraints,
\begin{equation} \label{eq:clusterstab}
    K_{x,y}|C\rangle=|C\rangle,
\end{equation}
where $K_{x,y}=X_{x,y} Z_{x+1,y} Z_{x-1,y} Z_{x,y+1} Z_{x,y-1}$ and $X_{x,y}$ ($Z_{x,y},Y_{x,y}$) denotes the Pauli operator $\sigma^X$ ($\sigma^Z,\sigma^Y$) acting on site $(x,y)$, see Fig.~\ref{fig:symms}. This means that the cluster state is a stabilizer state. It is therefore the unique ground state of the following Hamiltonian:
\begin{equation} \label{eq:clusterham}
    \mathcal{H}_C=-\sum_{x,y} K_{x,y}.
\end{equation}
This Hamiltonian commutes with subsystem symmetries consisting of $X$ acting on every site along any diagonal line on the lattice (corresponding to vertical and horizontal lines along the cylinder):
\begin{align} 
    U_v(c)&=\prod_{x=1}^N X_{x,c-x} \nonumber \\
    U_h(c)&=\prod_{x=-\infty}^\infty X_{x,c+x} \label{eq:symms}
\end{align}
We set our origin such that $U_v(0)$ corresponds to the symmetry lying parallel to the boundary of $A$, see Fig.~\ref{fig:symms}. The cluster state has SSPT order with respect to these symmetries \cite{Raussendorf2019,You2018,Devakul2018b}, and we call the corresponding SSPT phase of matter the cluster phase \cite{Raussendorf2019}. 

This paper deals with calculating the entropy of entanglement of the reduced density matrix $\rho_A$, where the $A$ subsystem corresponds to the right half of the cylinder. For much of this paper we will make statements about the structure of $\rho_A$ directly, such that our claims hold for any $\alpha$-R\'enyi entropy 
\begin{equation} \label{eq:Renyi}
  S_A^{(\alpha)}\equiv S^{(\alpha)}(\rho_A)=\frac{1}{1-\alpha}\log_2\mathrm{Tr}(\rho_A^\alpha),
\end{equation}
including the Von Neumann entropy $S^{(1)}_A=-\mathrm{Tr}(\rho_A \log_2 \rho_A)$ obtained in the limit $\alpha\rightarrow 1$. However, we will sometimes focus on $S_A^{(2)}$ as it is most amicable to our numerical methods. It is also the most convenient to measure experimentally \cite{Abanin2012,Daley2012,Islam2015}.
Whenever we make a statement that holds for all $\alpha$, we will simply use the notation $S$ or $S_A\equiv S(\rho_A)$.

As the cluster state is a stabilizer state, $S_A$ may be straightforwardly calculated. Let $G$ be the group generated by all stabilizers $K_{x,y}$, and let $G_A\subset G$ be the group of all elements of $G$ which act non-trivially only on region $A$. Then we have the following equation \cite{Hamma2005}:
\begin{equation} \label{eq:stabentropy}
    S_A=|A|-\log_2|G_A|.
\end{equation}
All stabilizers $K_{x,y}$ corresponding to lattice sites $(x,y)\in A - \partial A$ are contained in $G_A$. The product of all stabilizers along the boundary, which is precisely the line symmetry $U_{v}(0)$, is also in $G_A$, see Fig.~\ref{fig:symms}. Hence we have $|G_A|=2^{|A|-|\partial A|+1}$, so $S_A=N-1$. We see that the SPEE takes the value $\gamma=1$ $(=\log_2 2)$ for the cluster state, due to the subsystem symmetries forming non-local constraints lying along the boundary of $A$ \cite{Zou2016,Williamson2019,Schmitz2019}.

As a first venture away from the cluster state, we consider a family of states $|C(\theta)\rangle=U(\theta)|C\rangle$, where the circuit $U(\theta)$ is defined by acting with the two-body unitary $(H\otimes H) C\theta (H\otimes H)$ on every pair of neighbouring sites, with
\begin{equation} \label{eq:hadamard}
    H=\sqrt{\frac{1}{2}}\begin{pmatrix*}[r] 1 & 1 \\ 1 & -1 \end{pmatrix*},
\end{equation}
and $C\theta=\mathrm{diag}(1,1,1,e^{i\theta})$. $U(\theta)$ is diagonal in the local $X$-basis, hence it commutes with the subsystem symmetries of the cluster state, so $|C(\theta)\rangle$ is in the cluster phase for all $\theta$. We choose this family since it provides a smooth interpolation between the cluster state, $|C(0)\rangle$, and the state considered in Ref.~\cite{Williamson2019}, $|C(\pi)\rangle$, which was shown to display an enlarged value of $\gamma$.

For general $\theta$, $|C(\theta)\rangle$ is not a stabilizer state, so we need a different method to calculate its entropy. In order to calculate $S_A^{(2)}(\theta)$, we use the method of Ref.~\cite{Zou2016}. Notice that $|C(\theta)\rangle$ may be created by a unitary circuit acting on a product state. Therefore, by applying unitaries on the $A$ and $B$ subsystems separately, which does not change the entropy, we may disentangle all qubits except those on a strip along the boundary. Thus the calculation of entropy for our 2D system is reduced to that of a 1D system with an extensive bipartitioning, and this may be easily computed using a transfer matrix method. Namely, we can construct a matrix $Q(\theta)$ that follows from the definition of $|C(\theta)\rangle$, such that $S_A^{(2)}(\theta)=-\log_2 \mathrm{Tr}(Q(\theta)^N)$, see Ref.~\cite{Zou2016} for more details.
Let $\{\lambda_k\}_k$ be the set of eigenvalues of $Q(\theta)$ with maximum magnitude and write $\lambda_k=re^{i\phi_k}$. The entropy is then,
\begin{equation} \label{eq:tmentropy}
S_A^{(2)}(\theta)=N\log_2 r-\log_2 m+\dots,
\end{equation}
where $m=\sum_k e^{iN\phi_k}$ and the terms contained in the dots decay exponentially in $N$. We can identify the constants $a$ and $\gamma$ from Eq.~(\ref{eq:entropy}) with $\log_2 r$ and $\log_2 m$, respectively. Thus, the SPEE can be determined by examining the eigenvalues of $Q(\theta)$ with largest magnitude. If there are no non-positive eigenvalues with magnitude $r$, then the constant $m$ is simply the degeneracy of the largest eigenvalue. If such eigenvalues do exist, then $m$ can exhibit periodic $N$ dependence.

\begin{figure}
\centering
\includegraphics[width=\linewidth]{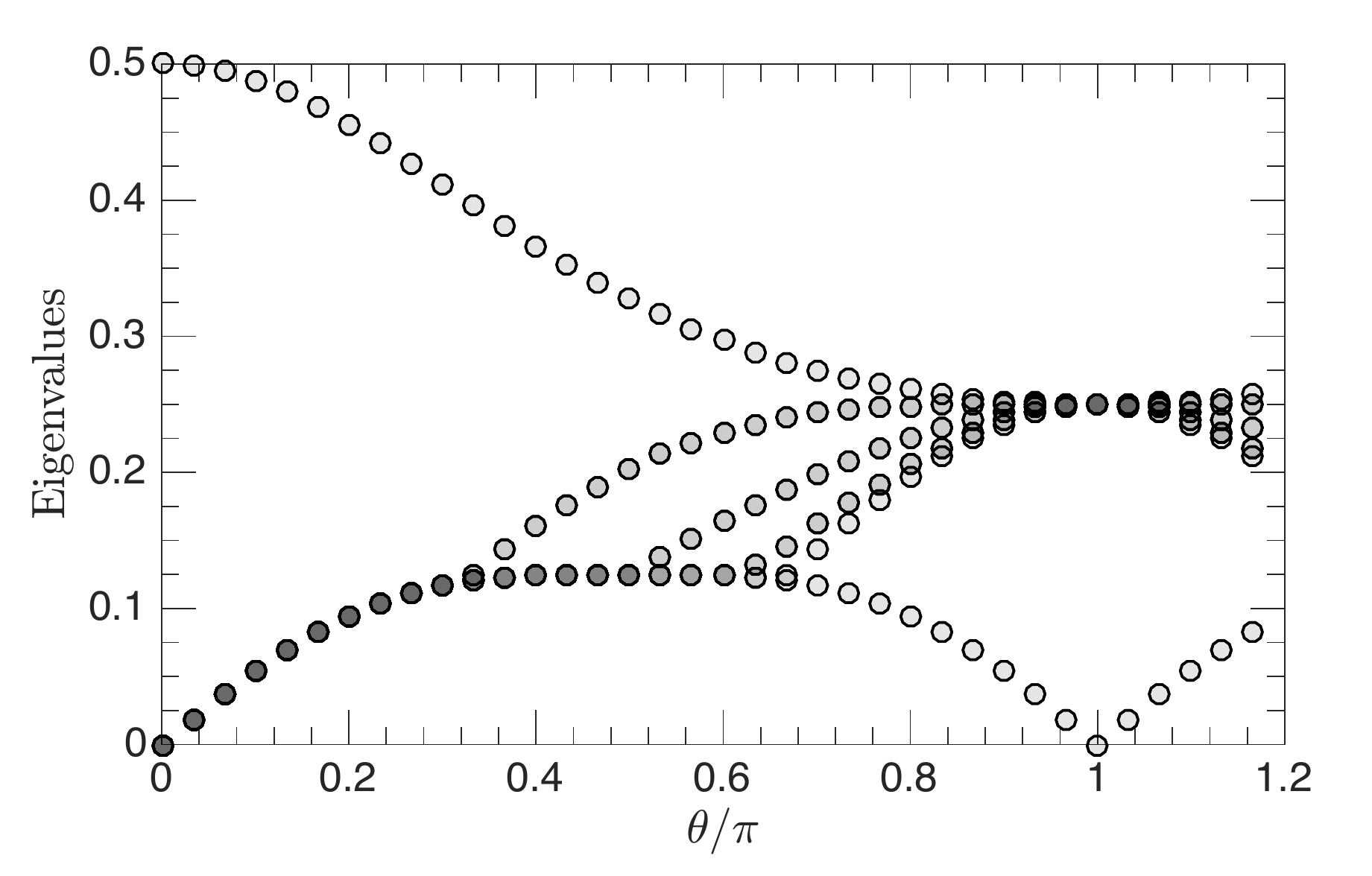} 
\caption{Spectrum of the transfer matrix $Q(\theta)$. For each value of $\theta$, the moduli of the 18 largest eigenvalues of $Q(\theta)$ are depicted, where each dot indicates at least 2 degenerate eigenvalues. At each value of $\theta$, the largest eigenvalue is two-fold degenerate, except for the point $\theta=\pi$ where it is 16-fold degenerate in magnitude.}
\label{fig:spectrum}
\end{figure}

The spectrum of $Q(\theta)$ can be computed exactly \footnote{This is possible because $Q(\theta)$ is simply a 256 $\times$ 256 matrix, which can be exactly diagonalized numerically}, and the results are shown in Fig.~\ref{fig:spectrum}. For $\theta\neq \pi$, there are two eigenvalues of largest magnitude, both of which are positive. Hence we have $m=2$ and $\gamma=1$. At $\theta=\pi$, there are 16 eigenvalues of largest magnitude. More precisely, we find that the non-zero eigenvalues of $Q(\pi)$ are $\frac{1}{4}$, $-\frac{1}{4}$, $\frac{i}{4}$, and $-\frac{i}{4}$ with degeneracies 8, 4, 2 and 2, respectively. In this case, we have $m=\left(8+4(-1)^N+2i^N+2(-i)^N\right)$, which gives,
\begin{equation} \label{eq:pientropy}
S_A^{(2)}(\pi)=\begin{cases} 2N-4 & \mbox{if } 4\mid N \\ 2N-3 & \mbox{if } 2\mid N\  \mbox{and } 4\nmid N \\ 2N-2 & \mbox{if } 2\nmid N \end{cases}.
\end{equation}
Therefore, the states $|C(\theta)\rangle$ have a SPEE $\gamma=1$ for all $\theta\neq \pi$, while $\gamma>1$ for $\theta=\pi$. 

The enlarged SPEE for $\theta=\pi$ can be attributed to the fact that $|C(\pi)\rangle$ has many extra symmetries, aside from those in Eq.~(\ref{eq:symms}), which are not satisfied for $\theta\neq \pi$. Namely, $|C(\pi)\rangle$ is a stabilizer state satisfying $K'_{x,y}|C(\pi)\rangle=|C(\pi)\rangle$ where the stabilizers $K'_{x,y}$ are as pictured in Fig.~\ref{fig:symms}. One can alternatively compute $S_A^{(2)}(\pi)$ using Eq.~(\ref{eq:stabentropy}), and the results agree with Eq.~(\ref{eq:pientropy}) \footnote{The enlarged SPEE comes from extra subsystem symmetries which appear in $G_A$. It would be interesting to investigate whether or not the enlarged SPEE would persist if these extra symmetries where preserved.}. At the end of the next section, we will use the tensor network representation of $|C(\pi)\rangle$ to more clearly identify the mechanism through which these extra symmetries lead to a larger SPEE. 

\section{Analytical argument for uniformity of the SPEE} \label{sec:anal}

The results of the previous section suggest that the SPEE $\gamma$ is uniform throughout the cluster phase, except for certain fine-tuned states with enhanced symmetries. Now, we will use tensor networks to give an analytical argument that all generic states in the cluster phase do indeed have the same SPEE. In fact, an elementary argument based purely on symmetry shows that the entanglement spectrum is $2^{N-1}$-fold degenerate, which implies that $S_A\geq N-1$, as described briefly in Appendix~\ref{app:sspt}. However, this does not tell us anything about the value of $\gamma$. The fact that $\gamma\geq 1$ was argued in Ref.~\cite{Devakul2018b} (using the result of Ref.~\cite{Zou2016}) under a similar set of assumptions as those used here \footnote{Namely, they assumed that, throughout the phase, we can disentangle the degrees of freedom far from the cut using unitary operators localized on either side of it. This is essentially the same as assuming a finite bond-dimension PEPS representation, as we do here.}.
The purpose of the more careful argument of this section is to argue that $\gamma=1$ exactly. We begin by reviewing the tensor network characterization of the cluster phase given in Ref.~\cite{Raussendorf2019}, which forms the basis of our argument, and then we show how this can be used to constrain $\gamma$. 

\subsection{Tensor network description of the cluster phase}

\begin{figure}
    \centering
    \includegraphics[width=0.7\linewidth]{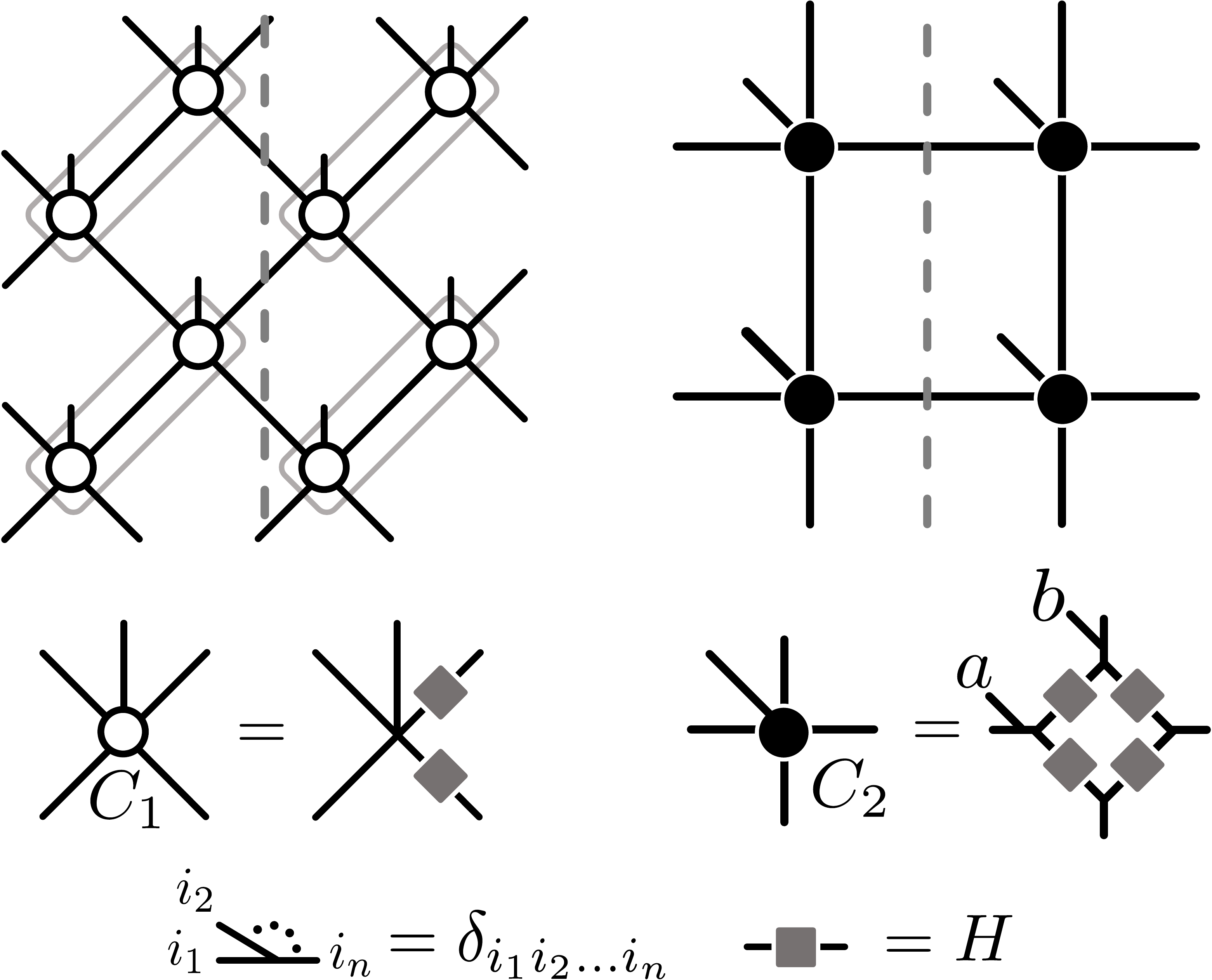}
    \caption{Two different tensor network representations of the 2D cluster state. On the left, we define the tensor $C_1$ which has the geometry of the rotated square lattice. On the right, we define the tensor $C_2$ which has two physical legs per tensor, labelled $a$ and $b$. The corresponding tensor network has the geometry of a (non-rotated) square lattice. Both tensors are defined in terms of Kronecker-$\delta$ tensors and the Hadamard matrix $H$ (Eq.~(\ref{eq:hadamard})). The dashed lines indicate the boundary between the $A$ and $B$ subsystems.}
    \label{fig:tn}
\end{figure}

In Fig.~\ref{fig:tn}, we define two PEPS representations of the cluster state. The first tensor, $C_1$, is the usual PEPS representation of the cluster state on the rotated square lattice. In Ref.~\cite{Raussendorf2019}, it was shown that every state in the cluster phase admits a PEPS representation on this lattice of the form,
\begin{equation} \label{eq:tensordecomp1}
    \includegraphics[scale=0.23]{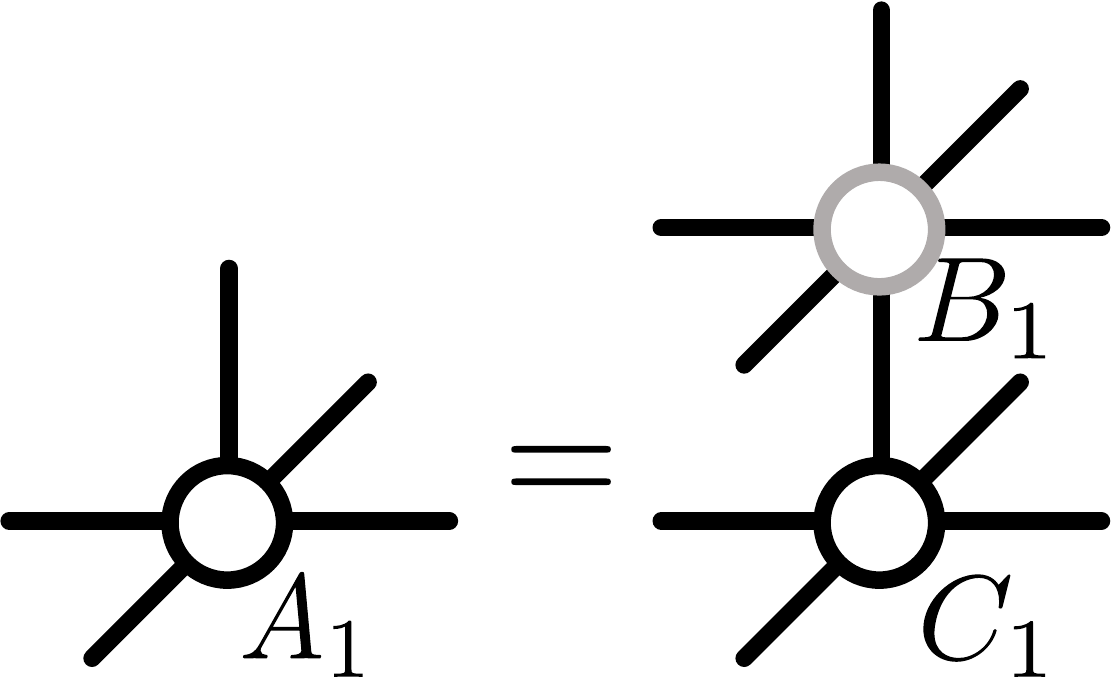}
    \quad \raisebox{0.4cm}{where} \quad \includegraphics[scale=0.23]{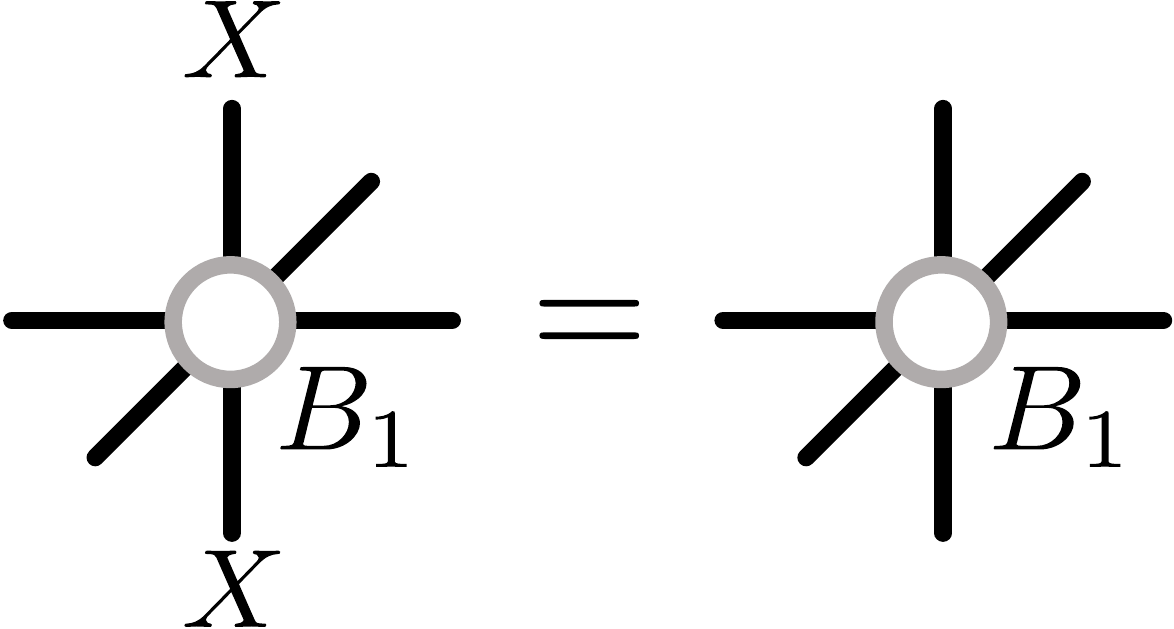}
\end{equation}
where $B_1$ is some tensor describing a projected entangled pair operator which commutes locally with $X$ as indicated. Throughout the cluster phase, $C_1$ stays the same, but $B_1$ varies \footnote{We note that the construction in Ref.~\cite{Raussendorf2019} gives PEPS tensors with exponentially large bond dimension, but this can be improved to give constant bond dimension when one takes advantage of the finite depth of the quantum circuits considered within.}. In this section, we will find it more convenient to consider a different PEPS representation of the cluster state, denoted $C_2$, which has two qubits per unit cell and is defined on a non-rotated square lattice, see Fig.~\ref{fig:tn}. Then the statement corresponding to Eq.~(\ref{eq:tensordecomp1}) is,
\begin{equation} \label{eq:tensordecomp2}
    \includegraphics[scale=0.23]{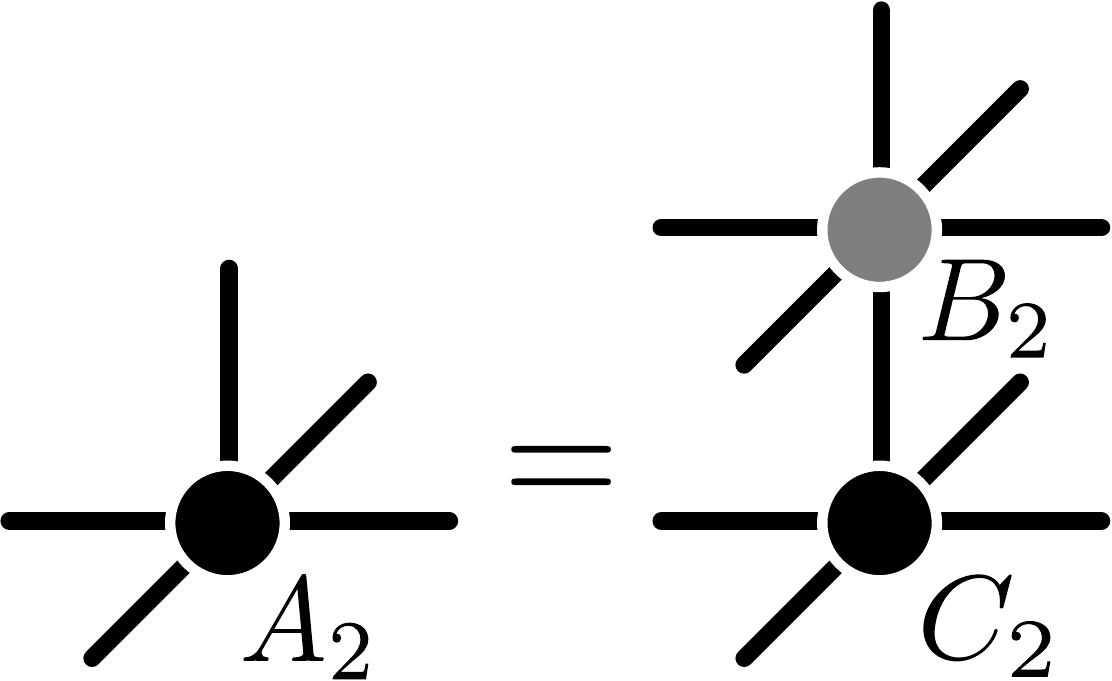}
    \quad \raisebox{0.4cm}{where} \quad \includegraphics[scale=0.23]{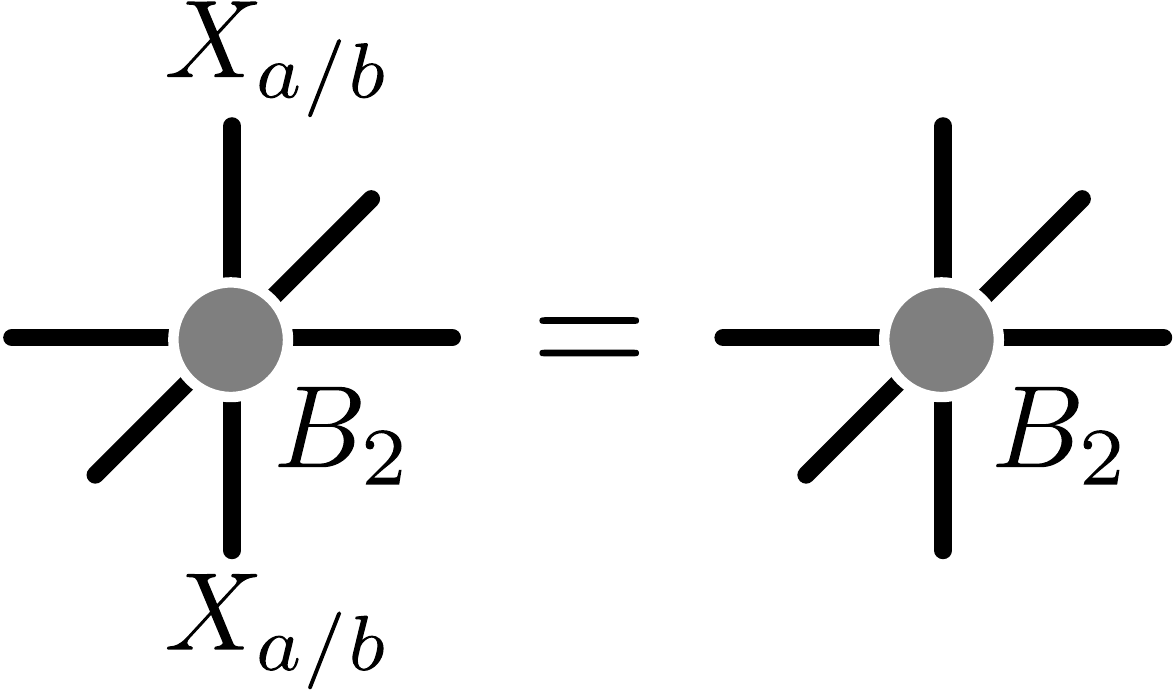}
\end{equation}
$B_2$ is again some tensor that varies through the phase. $X_{a/b}$ denotes $X$ acting on one of the two qubits in a unit cell, labelled $a/b$ as in Fig.~\ref{fig:tn}. 

If we contract the physical leg of any tensor $A$ with some state $|i\rangle\in \{|++\rangle ,|+-\rangle ,|-+\rangle, |--\rangle\}$, we obtain a tensor with 4 virtual legs, denoted $A^i$. In terms of these components, Eq.~(\ref{eq:tensordecomp2}) reads,
\begin{equation} \label{eq:tensordecomp3}
    A_2^i=B_2^i\otimes C_2^i\, .
\end{equation}
Thus, the virtual space of the tensor network decomposes into two subspaces, which are called the \textit{junk} and \textit{protected} subspaces, corresponding to the operators $B_2^i$ and $C_2^i$, respectively \cite{Else2012}. This decomposition is central to our arguments. It shows that the global pattern of entanglement that characterizes the cluster phase is that of the cluster state. An analogous statement is known to be true in certain 1D SPT phases, including that which contains the 1D cluster state \cite{Else2012}.

We now map our system onto a quasi-1D system by combining $2N$ spins around the cylinder into one larger block as shown in Fig.~\ref{fig:symms}. The associated tensor $\mathcal{A}$ is obtained by combining $N$ tensors $A_2$ in a ring around the cylinder, see Fig.~\ref{fig:tm}. If we let $\mathbf{i}=(i_1,i_2,\dots,i_N)$ be an element of the index set $I$ labelling all $2^{2N}$ states in a block, we can again define the tensor components $\mathcal{A}^\mathbf{i}$. Due to the subsystem symmetries $U_v(c)$, the wavefunction of our state consists only of terms with an even number of $|-\rangle$ states in both columns in each block. Denoting this even-parity subset of states $I_e\subset I$ and the corresponding odd-parity subspace as $I_o\subset I$, we can therefore modify the tensor $\mathcal{A}$ by setting $\mathcal{A}^\mathbf{i}=0$, $\forall \mathbf{i}\in I_o$, without changing the state described by $\mathcal{A}$. This modification simplifies our description later on.

For the remaining $\mathbf{i}\in I_e$, Eq.~(\ref{eq:tensordecomp3}) implies the decomposition $\mathcal{A}^\mathbf{i}=\mathcal{B}^\mathbf{i}\otimes \mathcal{C}^\mathbf{i}$, where $\mathcal{B}$ and $\mathcal{C}$ denote the blocked tensors living in the junk and protected subspaces, respectively. We can straightforwardly determine $\mathcal{C}^\mathbf{i}$ using the definition of $C_2$. We find that $\mathcal{C}^\mathbf{i}=\mathcal{P}^\mathbf{i}\Pi$, $\forall \mathbf{i}\in I_e$, where each $\mathcal{P}^\mathbf{i}$ is some tensor product of Pauli operators and
\begin{equation} \label{eq:projector}
    \Pi=\frac{\mathbb{1}+X^{\otimes N}}{2}
\end{equation}
is a rank-$2^{N-1}$ projection matrix with $X^{\otimes N}\equiv \prod_{k=1}^N X_k$. Therein, $X_k$ denotes $X$ acting on the $k$-th component of the $N$-component protected subspace. Importantly, the operators $\mathcal{P}^\mbf{i}$ also satisfy $\mathcal{P}^\mathbf{i}\Pi =\Pi \mathcal{P}^\mathbf{i}$, for all $\mbf{i}\in I_e$.

In Appendix~\ref{app:sspt}, we show how the non-trivial SSPT order of the cluster phase can be understood in terms of symmetry fractionalization on the virtual boundary of the PEPS. This follows directly from Eq.~(\ref{eq:tensordecomp2}), and is the PEPS equivalent of the picture developed in Ref.~\cite{Devakul2018b}.

\subsection{Constraining $\gamma$ in the cluster phase}

\begin{figure}
    \centering
    \includegraphics[width=\linewidth]{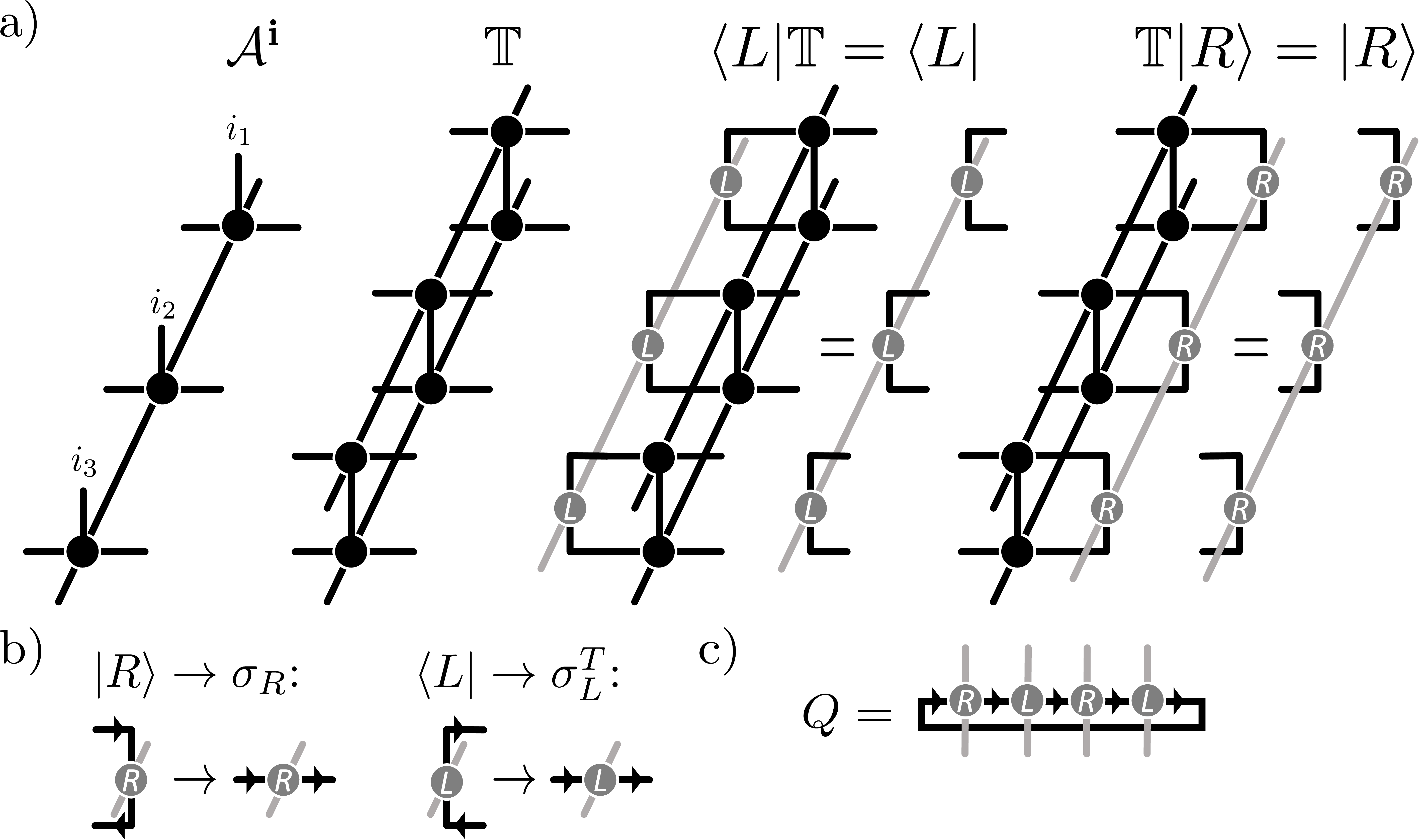}
    \caption{a) Illustration of the PEPS transfer matrix and its fixed points. b) Relation between the fixed points of $\mathbb{T}$ and $\mathcal{T}$ via reorientation of the legs of each tensor. c) The matrix $Q$ whose spectrum encodes the entanglement entropy for all system sizes $N$.}
    \label{fig:tm}
\end{figure}

We will now use the above characterization of the cluster phase to show that the entanglement entropy $S_A$ can be decomposed into two parts, $S_A=S_B+S_C$, where $S_C$ is the entropy of the cluster state, and $S_B$ is essentially the entropy of the PEPS defined by the tensor $B^i_2$. Since we know $S_C=N-1$, and we will argue $S_B$ satisfies an area law with no correction for generic $B^i_2$, we will find that $S_A$ has a correction of $\gamma=1$ for generic states in the cluster phase. 

First, we review the standard tensor network technique for calculating entanglement entropy \cite{Cirac2011}. We define the transfer matrix 
\begin{equation}
\mbb{T}=\sum_{\mbf{i}\in I} \mcal{A}^{\mbf{i}} \otimes \mcal{A}^{\mbf{i}*},
\end{equation}
where the star denotes complex conjugation. We can normalize our PEPS such that the largest eigenvalue of $\mbb{T}$ is 1, and denote by $|R\rangle$ and $\langle L|$ the left and right fixed-points of $\mbb{T}$, such that $\mbb{T}|R\rangle=|R\rangle$ and $\langle L|\mbb{T}=\langle L|$. For the following proof, we will find it more convenient to work with the associated quantum channel 
\begin{equation} \label{eq:tm}
    \mcal{T}(\rho)=\sum_{\mathbf{i}\in I} \mathcal{A}^\mathbf{i}\rho\mathcal{A}^{\mathbf{i}\dagger}
\end{equation}
which is related to $\mbb{T}$ by redefining the input and output legs of the tensor. We will refer to both $\mbb{T}$ and $\mcal{T}$ as the transfer matrix. The fixed points $|R\rangle$ and $\langle L|$ of $\mbb{T}$ correspond to fixed points $\sigma_R$ and $\sigma_L$ of $\mcal{T}$ and its adjoint $\mcal{T}^\dagger=\sum_{\mathbf{i}} \mathcal{A}^{\mathbf{i}\dagger}\rho\mathcal{A}^{\mathbf{i}}$, respectively. See Fig.~\ref{fig:tm} for a graphical representation of these objects.

An important property of the cluster phase is that the fixed point of the transfer matrix is unique, with all other eigenvalues having magnitude less than unity. Specifically, a degenerate fixed point space of the transfer operator would manifest itself as long-range correlations along the cylinder axis -- either for local operators, amounting to conventional long-range order~\cite{Rispler2015}, or for loop operators acting around the cylinder, amounting to topological order~\cite{Schuch2013}. Since any such order is absent in the cluster state (whose transfer operator has a unique fixed point), and thus by definition in the whole cluster phase, the fixed point of the transfer matrix must be unique in the whole cluster phase.

The reduced density matrix $\rho_A$ corresponding to half of the cylinder can be determined via a bulk-boundary correspondence for PEPS \cite{Cirac2011}. Namely, we can write $\rho_A=V\sigma V^\dagger$ where $V$ is an isometry and  $\sigma=\sqrt{\sigma_L^T}\sigma_R\sqrt{\sigma_L^T}$, with $T$ denoting the transpose. Thus, $S_A=S(\sigma)$. The rest of this section is dedicated to constraining $\sigma$ within the cluster phase.

To begin, note that, due to Eq.~(\ref{eq:tensordecomp3}), $\mathcal{T}$ satisfies many symmetries throughout the entire cluster phase. First, we have,
\begin{align}
&\mathcal{T}(\left[\mathbb{1}\otimes X_k\right] \rho \left[\mathbb{1}\otimes X_k\right]) \nonumber \\
&=\sum_{\mathbf{i}\in I} \mathcal{A}^\mathbf{i}\left[\mathbb{1}\otimes X_k\right] \rho \left[\mathbb{1}\otimes X_k\right]\mathcal{A}^{\mathbf{i}\dagger} \nonumber \\
&=\sum_{\mbf{i}\in I_e} \left(\mcal{B}^{\mbf{i}}\otimes \mcal{P}^{\mbf{i}}\Pi X_k\right) \rho \left(\mcal{B}^{\mbf{i}} \otimes X_k \Pi\mcal{P}^{\mbf{i}} \right) \nonumber \\
&=\sum_{\mbf{i}\in I_e} \left(\mcal{B}^{\mbf{i}}\otimes X_k \mcal{P}^{\mbf{i}}\Pi\right) \rho \left(\mcal{B}^{\mbf{i}} \otimes\Pi\mcal{P}^{\mbf{i}} X_k \right) \nonumber \\
&=\left[\mathbb{1}\otimes X_k\right] \mathcal{T}( \rho ) \left[\mathbb{1}\otimes X_k\right]. \label{eq:tmsymm1}
\end{align}
Therein, $X_k$ acts on the protected subspace, while $\mathbb{1}$ acts on the junk subspace. In the second equality, we substituted $\mathcal{A}^\mathbf{i}=\mathcal{B}^\mathbf{i}\otimes \mathcal{P}^\mathbf{i}\Pi$. In the third equality, we used the facts that $X_k\mcal{P}^{\mbf{i}}=\pm\mcal{P}^{\mbf{i}}X_k$ and $X_k \Pi=\Pi X_k$. In the same way, we can derive,
\begin{equation} \label{eq:tmsymm2}
    \mcal{T}(Z_kZ_{k+1}\rho Z_kZ_{k+1})=Z_kZ_{k+1}\mcal{T}(\rho)Z_kZ_{k+1}.
\end{equation}
Above, and in the rest of this section, we have omitted the identity operators acting on the junk subspace for notational simplicity. Finally, because $X^{\otimes N}\Pi=\Pi$, we also have the symmetries,
\begin{equation} \label{eq:tmsymm3}
    \mcal{T}(\rho)=X^{\otimes N}\mcal{T}(\rho)=\mcal{T}(\rho)X^{\otimes N}. 
\end{equation}
All of the above symmetries hold also for $\mcal{T}^\dagger$. These symmetries are a reflection of the symmetry fractionalization on the boundary that characterizes the cluster phase, as described in Appendix~\ref{app:sspt}.

Since the fixed-points are unique, they inherit these symmetries. For example, if $\sigma_R$ is the fixed-point of $\mathcal{T}$, Eq.~(\ref{eq:tmsymm1}) implies that $X_k\sigma_R X_k$ is as well. Since the fixed-point is unique up to a multiplicative constant, we have $X_k\sigma_R X_k\propto \sigma_R$. Since $\sigma_R$ is positive, and conjugation by $X_k$ preserves the trace, we must have $X_k\sigma_R X_k=\sigma_R$. Using similar arguments, we get all of the following symmetries:
\begin{align}
\left[\sigma_{R/L}, X_k\right]&=\left[\sigma_{R/L}, Z_kZ_{k+1}\right]=0,  \nonumber \\
 X^{\otimes N} \sigma_{R/L}&=\sigma_{R/L}X^{\otimes N}=\sigma_{R/L}. 
\end{align}
These symmetries completely constrain $\sigma_{R/L}$ on the protected subspace, with $\Pi$ being the unique solution to the constraints. Thus, the fixed points decompose across the junk and protected subspaces as $\sigma_{R/L}=\widetilde{\sigma}_{R/L}\otimes \Pi$ for some unknown states $\widetilde{\sigma}_{R/L}$. This gives $\sigma=\widetilde{\sigma}\otimes \Pi$, where $\widetilde{\sigma}=\sqrt{\widetilde{\sigma}_L^T}\widetilde{\sigma}_R\sqrt{\widetilde{\sigma}_L^T}$. 

Having constrained the fixed points in the protected subspace, what can we now say about the unconstrained part $\widetilde{\sigma}$? It turns out that $\widetilde{\sigma}_{R/L}$ are themselves fixed points of a certain transfer operator. Indeed, we have,
\begin{align}
    \mathcal{T}(\widetilde{\sigma}_R\otimes \Pi)&=\sum_{\mathbf{i}\in I} \mathcal{A}^\mathbf{i}(\widetilde{\sigma}_R\otimes \Pi)\mathcal{A}^{\mathbf{i}\dagger} \nonumber \\
    &=\sum_{\mathbf{i}\in I} \mathcal{B}^\mathbf{i}\widetilde{\sigma}_R\mathcal{B}^{\mathbf{i}\dagger} \otimes \mathcal{C}^\mathbf{i}\Pi\mathcal{C}^{\mathbf{i}\dagger} \nonumber \\
    &=\sum_{\mathbf{i}\in I_e} \mathcal{B}^\mathbf{i}\widetilde{\sigma}_R\mathcal{B}^{\mathbf{i}\dagger}\otimes \Pi 
    \label{eq:factored-fixedpt}
\end{align}
where we used the facts that $\mcal{C}^\mbf{i}=0$, $\forall \mbf{i}\in I_o$, and $[ \mathcal{C}^\mathbf{i}, \Pi ]=0$, $\forall\mathbf{i}\in I_e$. Since $\mathcal{T}(\widetilde{\sigma}_R\otimes \Pi)=\widetilde{\sigma}_R\otimes \Pi$ by definition, this gives
\begin{equation} \label{eq:junktm}
    \widetilde{\mathcal{T}}(\widetilde{\sigma}_R):=\sum_{\mathbf{i}\in I_e} \mathcal{B}^\mathbf{i}\widetilde{\sigma}_R\mathcal{B}^{\mathbf{i}\dagger}= \widetilde{\sigma}_R\ .
\end{equation}
Furthermore, Eq.~\eqref{eq:factored-fixedpt} implies that every eigenvector $\widetilde{\rho}$ of $\widetilde{\mathcal{T}}$ yields an eigenvector $\widetilde{\rho}\otimes \Pi$ of $\mathcal{T}$ with the same eigenvalue. Therefore, $\widetilde{\sigma}_R$ is the unique fixed-point of $\widetilde{\mcal{T}}$, since the fixed-point of $\mathcal{T}$ is unique. Similarly, $\widetilde{\sigma}_L$ is the unique fixed-point of $\widetilde{\mathcal{T}}^\dagger$.

 Since $\sigma=\widetilde{\sigma}\otimes\Pi$, the entanglement entropy decomposes into two components,
\begin{equation}
S_A=S(\sigma)=S(\widetilde{\sigma})+S(\Pi)\ ,
\end{equation}
where $S(\Pi)=N-1$ is the entropy of the cluster state.  What remains is to understand the entropy contribution $S(\widetilde\sigma)$. 
We will argue that $S(\widetilde\sigma)$ generically satisfies an area law with no correction.
To this end, consider a generic tensor $A_2^i=B_2^i\otimes C_2^i$ which is in the cluster phase.  Therefore, it can be connected to the cluster state via a smooth path 
$A_2^i(\theta)=B_2^i(\theta)\otimes C_2^i$ without closing the gap of the transfer operator $\mcal{T}$. This implies that 
$B_2^i(\theta)$ smoothly connects $B_2^i$ to a trivial tensor, where the corresponding transfer operator $\widetilde{\mcal{T}}$ must remain gapped along the path as well.
Up to the restriction $\mathbf{i}\in I_e$, $\widetilde{\mathcal T}$ is thus the transfer operator of a system in the trivial phase.  However, we expect that such a global parity constraint will only affect the entropy if the system either has topological order -- which we have ruled out -- or physical symmetries in the basis of the constraint, which would require fine-tuning of the $B_2^i$. Thus, we expect the fixed point of $\widetilde{\mathcal T}$ to have the structure of a generic fixed point in the trivial phase, which does not exhibit long-range correlations and thus no constant correction to the area law. 

Overall, this reasoning implies that, for generic points in the cluster phase, the entropy $S(\widetilde\sigma)$ should scale as
\begin{equation}
S(\widetilde\sigma) \propto N
\end{equation}
with corrections vanishing as $N\to\infty$, whereas constant corrections are expected only at fine-tuned points with additional symmetries. 
This is confirmed by numerical study of generic tensors $B_2^i$ up to bond dimension $4$.
We thus find that, for a generic point in the cluster phase, the entanglement entropy should scale like $S_A = aN-1$ for some constant $a$.

In light of these arguments, let us now reconsider why $|C(\pi)\rangle$ from Sec.~\ref{sec:cluster} has a correction $\gamma>1$. For this state, one can write a PEPS in the form of Eq.~(\ref{eq:tensordecomp2}), where $B^i$ is the same tensor as $C^i$, up to applying $H$ on each physical leg. Therefore, in addition to Eqs.~(\ref{eq:tmsymm1}-\ref{eq:tmsymm3}), $\mathcal{T}$ has just as many extra symmetries which act non-trivially on the junk subspace. These symmetries serve to constrain $\widetilde{\sigma}_{R/L}$, leading to the increased value of $\gamma$. Such symmetries are generically not present in the cluster phase, and indeed disappear for all $\theta\neq\pi$, reflected by the generic correction $\gamma=1$.

\section{Numerical detection of SSPT order} \label{sec:num}

Above, we have argued that $\gamma=1$ within the entire cluster phase, except at certain fine-tuned points of enhanced symmetry. In this section, we perform numerical calculations of $\gamma$ in ground states. The motivation of this is twofold. First, we would like to substantiate our analytical arguments. By considering known models, we will confirm that $\gamma=1$ within the cluster phase, and $\gamma=0$ in the trivial phase. Second, once we are confident that $\gamma=1$ indicates the presence of SSPT order, we can use it as a probe for phase transitions. We will add various fields and interactions onto the cluster Hamiltonian, with phase transitions into the trivial phase indicated by $\gamma$ suddenly dropping from 1 to 0. This study will lead us to the discovery of a new phase of matter beyond the cluster phase.

\subsection{Description of the algorithm}

We use the following method to calculate $\gamma$ in ground states. Given a Hamiltonian $H$, we use infinite projected entangled pair state (iPEPS) techniques to obtain a PEPS approximation to the ground state \cite{Vanderstraeten2016}. We then use this PEPS to construct the transfer matrix as in Eq.~(\ref{eq:tm}), and use an infinite matrix product state (iMPS) algorithm to obtain matrix product operator (MPO) approximations for the fixed-points $\sigma_R$ and $\sigma_L$. Using this, we can write
\begin{equation} S_A^{(2)}=-\log_2\mathrm{Tr}(\sigma_R\sigma_L^T\sigma_R\sigma_L^T)=-\log_2\mathrm{Tr}\left(Q^N\right)\, ,
\end{equation} 
where $Q$ is a matrix obtained from the MPO tensors describing $\sigma_R$ and $\sigma_L^T$, see Fig.~\ref{fig:tm} \footnote{We remark that the matrix $Q$ is not the same object as the matrix $Q(\theta)$ introduced in Sec.~\ref{sec:cluster}, although they are closely related}. For all states encountered in this section, the leading eigenvalue of $Q$ will either be unique or two-fold degenerate. Denote the two largest eigenvalues of $Q$ (which may or may not be degenerate) as $\lambda_0$ and $\lambda_1$. In analogy with Eq.~(\ref{eq:tmentropy}), we have
\begin{equation} \label{eq:qentropy}
    S^{(2)}_A= N\log_2 \lambda_0- \gamma+\dots\, ,
\end{equation}
where $\gamma=1$ if $\lambda_0=\lambda_1$, and $\gamma=0$ otherwise. Therefore, we can determine $\gamma$ from the ratio $\lambda_1/\lambda_0$. We note that this ratio is closely related to the replica correlation length that was introduced in Ref.~\cite{Zou2016} to study spurious corrections to the area law. A similar method to compute entanglement entropy in PEPS was also proposed in Ref.~\cite{Poilblanc2016}.

\subsection{Stability of the cluster phase}

Now we use the above method to investigate the fate of the SPEE under various local fields or interactions added to the cluster Hamiltonian, one at a time. The goal is to support our argument that the value $\gamma=1$ characterizes the cluster phase, meaning that this value persists unless the subsystem symmetries are explicitly broken or a phase transition is crossed.

\begin{figure}
    \centering
    \includegraphics[width=\linewidth]{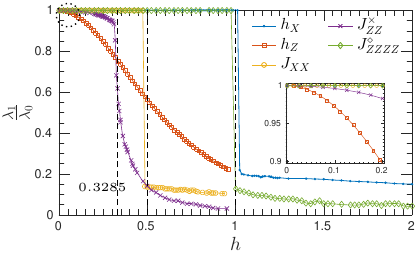}
    \caption{Entanglement entropy versus field or interaction strength of various terms added to the cluster Hamiltonian. $h$ is a dummy variable, standing for the different variables indicated in the legend. The y-axis shows the ratio of the two largest eigenvalues of the transfer matrix $Q$ defined in Fig.~\ref{fig:tm}. There is a correction $\gamma=1$ to the area law if $\lambda_1/\lambda_0=1$, and otherwise it is 0. The inset zooms in on the region indicated by the dotted circle, and shows that the symmetry-breaking terms destroy the correction for any finite value of $h$.}
    \label{fig:ee}
\end{figure}

We consider a Hamiltonian of the form:
\begin{equation} \label{eq:hams}
     \mathcal{H}=\mathcal{H}_C+\sum_{x,y} h'_{x,y}\, ,
\end{equation}
for several different choices of $h'_{x,y}$, each with different symmetries. In Fig.~\ref{fig:ee}, we plot $\lambda_1/\lambda_0$ as a function of the strength of the perturbation $h'_{x,y}$. The important findings are the following: 

(i) When a small symmetry respecting perturbation is added to the Hamiltonian, the SPEE keeps the value $\gamma=1$. 

(ii) When the strength of the symmetry respecting perturbations are increased, we encounter phase transitions indicated by a sudden drop to $\gamma=0$. The location of the phase transitions agree with known results where available. 

(iii) Adding a term which does not commute with the subsystem symmetries removes the SPEE for any finite coupling strength.

Overall, these results support our analytical results and confirm that the SPEE is a useful probe for detecting SSPT order and phase transitions. We will now discuss each choice of $h'_{x,y}$ in more detail. 

\subsubsection{Symmetry respecting terms}

We first consider adding terms which commute with the subsystem symmetries. For such terms, we expect that the value of the SPEE will remain constant up until an SSPT phase transition is reached, at which point it should should jump to 0. The first term we consider is a simple local $X$-field:
\begin{equation} \label{eq:hamx}
h'_{x,y}=-h_X X_{x,y}\, .
\end{equation}
The model of a cluster Hamiltonian with added $X$-field been studied previously in Refs.~\cite{Doherty2009, Kalis2012, Orus2013, You2018}. For $h_X\rightarrow\infty$, the model becomes a trivial paramagnet without SSPT order, so there must be a phase transition at some value of $h_X$. The model is self dual under the unitary circuit $U_{CZ}:=U(\pi)$ (see Sec.~\ref{sec:cluster}), so this phase transition should occur at $h_X=1$. Indeed, via a non-local duality transformation, this model can be mapped to the so-called Xu-Moore model with a transverse field, which is known to have a first order phase transition at $h_X=1$ \cite{Xu2004,Doherty2009, You2018}. In agreement with these facts, we find that $\lambda_1/\lambda_0$ changes from 1 discontinuously at the transition point, such that $\gamma$ jumps to 0 at this point, thereby correctly detecting the phase transition.

The next term we consider is a four-body interaction,
\begin{equation} \label{eq:hamzzzz}
h'_{x,y}=-J^\diamond_{ZZZZ} Z_{x-1,y}Z_{x+1,y}Z_{x,y-1}Z_{x,y+1}\, .
\end{equation}
This is the minimal term that contains $Z$ operators yet still commutes with all subsystem symmetries. Interestingly, this model behaves in the opposite way to first one, in that it is mapped to the Xu-Moore model under $U_{CZ}$, while it is self-dual under the same non-local duality transformation. These facts predict a phase transition at $J^\diamond_{ZZZZ}=1$, and this is again in agreement with the behaviour of $\lambda_1/\lambda_0$.

The final symmetry-preserving term we consider is a nearest-neighbour Ising interaction,
\begin{equation} \label{eq:hamxx}
h'_{x,y}=-J_{XX} \left( X_{x,y}X_{x+1,y}+X_{x,y}X_{x,y+1}\right) .
\end{equation}
To the best of our knowledge, this model has yet to be studied in the literature. Via the same non-local duality transformation mentioned earlier, it may be mapped onto two Xu-Moore models coupled by Ising interactions. The SPEE disappears at $J_{XX}=0.5$, suggesting that there is an SSPT phase transition into the symmetry-breaking phase at this point. 

We also verified that simultaneously adding all three symmetry respecting terms with random small couplings does not change the SPEE. This shows that the behaviour observed above is not a consequence of the specific Hamiltonians considered, and is instead generic behaviour.

\subsubsection{Symmetry breaking terms}

We now turn our attention to terms which anti-commute with the subsystem symmetries, and therefore explicitly break them. For these terms, the SSPT order should be destroyed for any finite coupling strength, so $\lambda_1/\lambda_0$ should decrease from 1 immediately. The simplest symmetry-breaking term is a $Z$-field,
\begin{equation} \label{eq:hamz}
h'_{x,y}=-h_Z  Z_{x,y}.
\end{equation}
As predicted, we find $\lambda_1/\lambda_0<1$ for any value of $h_Z$. Furthermore, this model is mapped to a trivial Hamiltonian under the action of $U_{CZ}$. Therefore, there is no phase transition for any value of $h_Z$, and this is consistent with the smooth decay of $\lambda_1/\lambda_0$ that we observe.

The next term we consider is a next-nearest neighbour interaction,
\begin{equation} \label{eq:hamzz}
h'_{x,y}=-J^\times_{ZZ}\left(Z_{x,y}Z_{x+1,y+1}+Z_{x,y}Z_{x-1,y+1}\right).
\end{equation}
While this term anti-commutes with the subsystem symmetries, it commutes with the global ``checkerboard'' $\mathbb{Z}_2\times\mathbb{Z}_2$ symmetry of the cluster state, which is generated by applying $X$ to all even or all odd spins on the square lattice. The cluster state has ``weak'' 2D SPT order under this global symmetry group, which is defined by a 2D SPT order that is only non-trivial in the presence of translational invariance \cite{Chen2013,Miller2016}. Therefore, it is important to confirm that the non-zero value of the SPEE is due to the subsystem symmetries, and not the global symmetries alone. Indeed, we find $\lambda_1/\lambda_0<1$ for any finite value of $J^\times_{ZZ}$, indicating that the global symmetries are not sufficient to protect the SPEE. As opposed to the previous case, we observe some singular behaviour of $\lambda_1/\lambda_0$ as $J^\times_{ZZ}$ is increased. This is explained by the fact that, under $U_{CZ}$, this model is mapped to two decoupled 2D quantum Ising models, which undergo a second order phase transition into a symmetry-breaking phase at $J^\times_{ZZ}\approx 0.3285$. Thus, the behaviour of $\lambda_1/\lambda_0$ also helps to detect non-SSPT phase transitions.

\subsection{Beyond the cluster phase --- Time reversal symmetry} \label{sec:xy}

\begin{figure}
    \centering
    \includegraphics[width=\linewidth]{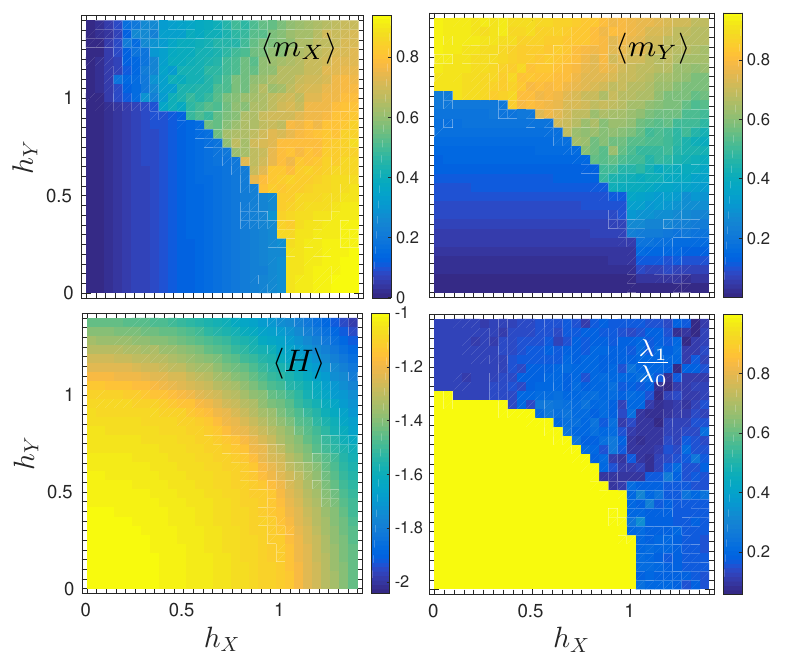}
    \caption{Phase diagram of the cluster model with added local Pauli $X$ and $Y$ fields with field strengths $h_X$ and $h_Y$, respectively. The local magnetizations $\langle m_X\rangle$ and $\langle m_Y\rangle$ along the $X$ and $Y$ directions, and the ground state energy per site $\langle H\rangle$ are plotted alongside the ratio $\lambda_1/\lambda_0$ of the two largest eigenvalues of the matrix $Q$. There is a large region in which this ratio is very nearly 1, indicating that the SPEE takes the value $\gamma=1$ in this region.}
    \label{fig:2d}
\end{figure}

We now consider adding the term,
\begin{equation} \label{eq:hamy}
    h'_{x,y}=-h_Y Y_{x,y}.
\end{equation}
This term anti-commutes with the subsystem symmetries, as with Eq.~\eqref{eq:hamz}. Therefore, according to our findings in the previous subsection, we would expect any finite value of $h_Y$ to remove the SPEE. This turns out not to be the case, as seen in Fig.~\ref{fig:2d}, and this section is devoted to understanding this behaviour.

It was shown in Ref.~\cite{Stephen2019} that the cluster state also has non-trivial SSPT order protected by fractal subsystem symmetries which are composed of tensor products of $Y$ operators. Since the model is trivial when $h_Y\rightarrow\infty$, there should be a phase transition into the trivial phase for some $h_Y$. Since the model is self-dual under $U_{CZ}$ followed by applying $S=\mathrm{diag}(1,i)$ on every site, this transition should occur at $h_Y=1$. Along the $y$-axes of Fig.~\ref{fig:2d}, we see good evidence that this is the case. For example, the magnetization along the $Y$ axis jumps discontinuously at this point.

What is surprising is that this transition is also detected by the SPEE, which remains equal to 1 up until $h_Y=1$, even though the $Y$-field does not commute with the subsystem symmetries. It is tempting to attribute the SPEE to the fractal symmetries of the cluster state, which do commute with the $Y$-field, but this is not the case. Consider simultaneously adding $X$ and $Y$ magnetic fields to the cluster state, 
\begin{equation} \label{eq:hamxy}
h'_{x,y}=-h_X X_{x,y}-h_Y Y_{x,y}.
\end{equation}
We see in Fig.~\ref{fig:2d} that the SPEE still persists in a large region. This cannot be due to the $X$ linelike symmetries of Eq.~\eqref{eq:symms}, nor the $Y$ fractal symmetries, since both are explicitly broken by the added fields. Therefore, it appears that we have discovered a new SSPT phase which contains at least part of cluster phase, and which is also accompanied by a non-zero SPEE. 

A similar phenomenon occurs in 1D. The 1D cluster state has SPT order under a global $\mathbb{Z}_2\times\mathbb{Z}_2$ symmetry, generated by applying $X$ to all even or all odd spins on the chain. Similar to the current scenario, this SPT order is stable under adding a $Y$-field, despite the fact that this is a symmetry-breaking term. In this case, the resolution is that the 1D cluster state also has time-reversal symmetry \cite{Verresen2017}. This symmetry also protects the SPT order of the cluster state, and commutes with the $Y$-field. 

Time-reversal turns out to be the solution here as well, although it takes an unusual form. In Ref.~\cite{You2018}, the authors defined subsystem time-reversal symmetries as subsystem unitary symmetries, such as the linelike symmetries considered here, followed by \textit{global} time-reversal. But it is easy to see that this also does not work, since the global time-reversal flips the sign of all $Y$-fields, not just those lying along a given line of subsystem symmetry. We therefore need a different notion of subsystem time-reversal, in which we enact time-reversal \textit{locally} only on those sites of the lattice on which the subsystem symmetry acts non-trivially.

Ref.~\cite{Chen2015} suggested one way to implement time-reversal locally on a tensor network. Globally, time-reversal symmetry acts on a quantum state by the action of a unitary operator (here, the subsystem symmetries) combined with complex conjugation of the wavefunction. With tensor networks, the wavefunction is divided into local tensors. Thus, one can define local complex conjugation at a given site by conjugating only the tensor associated to that site. More precisely, we can define operators $\mcal{K}_{x,y}$ which act on the PEPS by conjugating the tensor at site $(x,y)$ only. Note that, to ensure that the proper notion of locality is used, we use the tensor network defined on a rotated square lattice with a single-qubit unit cell, shown in Fig.~\ref{fig:tn}.

 With this, we define our subsystem time-reversal as,
\begin{align} 
    U^T_{v}(c)&=\prod_{x=1}^N X_{x,c-x}\mcal{K}_{x,c-x}, \nonumber \\
    U^T_{h}(c)&=\prod_{x=-\infty}^\infty X_{x,c+x}\mcal{K}_{x,c+x}. \label{eq:trsymms}
\end{align}
Is this the correct symmetry to describe the phase of matter observed in Fig.~\ref{fig:2d}? To answer this, we consider a state described by a tensor of the following form,
\begin{equation} \label{eq:perturbedA}
    \includegraphics[scale=0.23]{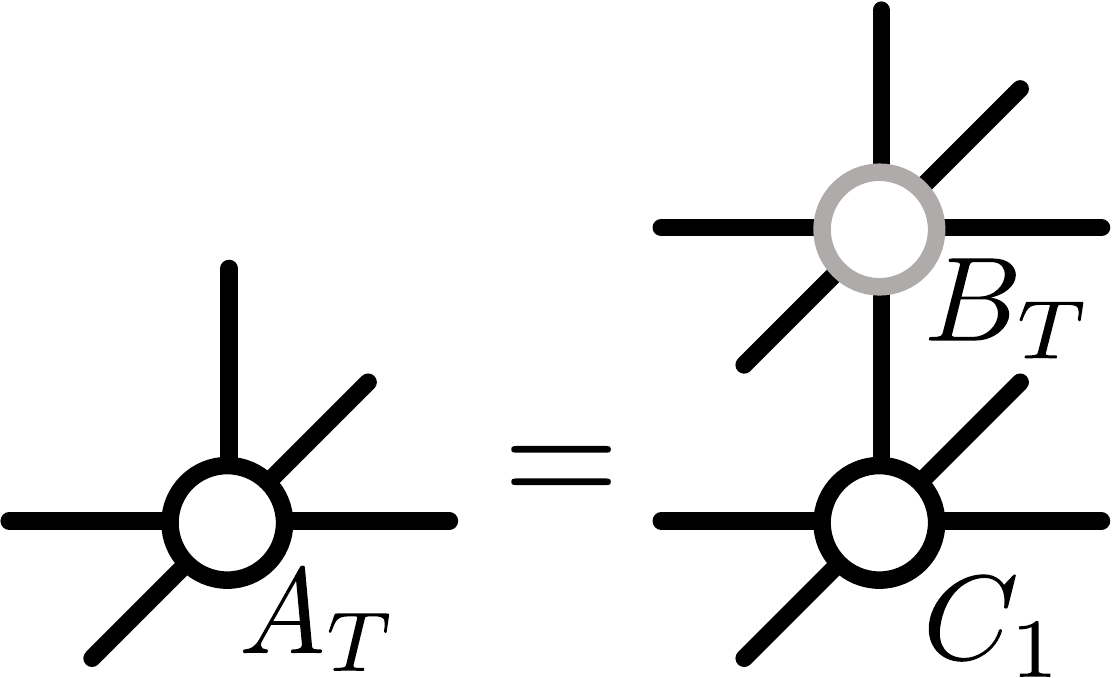}
    \quad \raisebox{0.4cm}{where} \quad \includegraphics[scale=0.23]{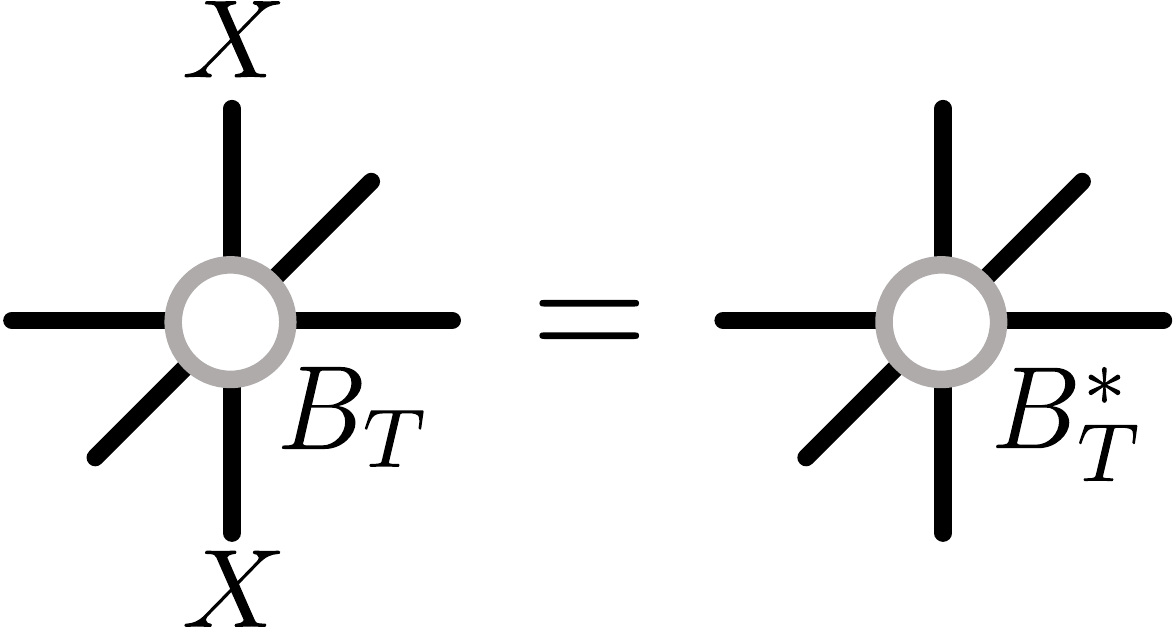}
\end{equation}
where $C_1$ is the cluster tensor, and $B_T^*$ is the complex conjugate of $B_T$. The state described by a tensor of this form is not generally symmetric under $U_{h,v}(c)$, but it is symmetric under $U^T_{h,v}(c)$ (note that the cluster state tensor $C_1$ is real, and hence unaffected by $\mcal{K}$). Furthermore, if we pick a random bond dimension 4 tensor $A_T$ of the form given in Eq.~(\ref{eq:perturbedA}), then we can confirm numerically that the resulting PEPS does indeed have $\gamma=1$. 

We conjecture that ground states in the non-trivial phase in Fig.~\ref{fig:2d} can be captured by tensors of the form Eq.~(\ref{eq:perturbedA}), in the same way that the cluster phase is captured by tensors of the form Eq.~(\ref{eq:tensordecomp1}). In Appendix~\ref{app:perturbation}, we argue that this is indeed the case by using the framework of perturbation theory in PEPS \cite{Vanderstraeten2017}. For example, to first order in perturbation theory, we can approximate the ground states of the model described by Eqs.~(\ref{eq:hams},\ref{eq:hamxy}) with the following PEPS tensor,
\begin{equation} \label{eq:clambda}
     \includegraphics[scale=0.23]{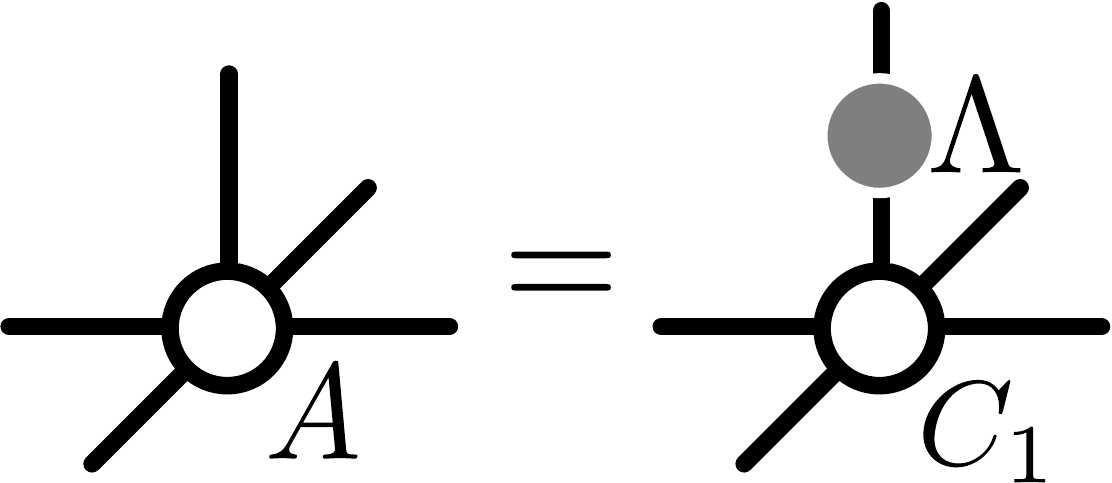}
\end{equation}
where the matrix $\Lambda$ is defined as,
\begin{equation} \label{eq:lambda}
    \Lambda=\mathbb{1}-\frac{h_X}{8}X-\frac{h_Y}{10}Y.
\end{equation}
Notice that $\Lambda$ satisfies $X\Lambda X=\Lambda^*$, such that $A$ is of the form given in Eq.~(\ref{eq:perturbedA}). Furthermore, one can confirm that the fixed-point $\sigma$ of the associated transfer matrix is exactly $\Pi$, independent of the values of $h_X$ of $h_Y$, so we have $\gamma=1$.

A similar argument holds to all orders in perturbation theory. Therefore, at least up to the phase transition where perturbation theory breaks down, the ground states in the non-trivial phase of Fig.~\ref{fig:2d} can be captured by PEPS of the form Eq.~(\ref{eq:perturbedA}), which we believe will generically have $\gamma=1$, using similar arguments as in Sec.~\ref{sec:anal}. On the other hand, if we also add a $Z$ field to the Hamiltonian, the tensor obtained from perturbation theory no longer has the form of Eq.~(\ref{eq:perturbedA}) and, accordingly, the SPEE disappears. 

Overall, our understanding of the phase of matter in Fig.~\ref{fig:2d} comes from three steps: (1) The subsystem time reversal symmetries $U^{T}_{h,v}(c)$ are the relevant symmetries that protect the SSPT order and (2) states in the same phase as the cluster state under $U^{T}_{h,v}(c)$ can be captured by PEPS of the form Eq.~(\ref{eq:perturbedA}), which (3) generically have a SPEE of $\gamma=1$. We leave a more rigorous confirmation of steps (1)-(3) to future work.

\subsection{Remarks on the numerical method}

We finish this section by commenting on some relevant features of our numerical method. To begin, we briefly compare our method of computing $\gamma$ to the more common approach, which involves calculating ground states and their respective entropies $S_A$ for several system sizes $N$, and then making a linear fit to extract $\gamma$ \cite{Jiang2012}. This method is relatively costly, in that it requires the determination of several ground states, and it is also sensitive to finite size effects. On the other hand, our method can obtain $\gamma$ from one PEPS tensor, and works directly in the thermodynamic limit, such that finite size effects are minimized.

Second, we see from Figs.~\ref{fig:ee} and \ref{fig:2d} that the SPEE can serve as a very good probe for SSPT phase transitions, with the phase boundary marked by a clear discontinuity. In some cases, the phase transitions observed here may also be detected by discontinuities in local magnetizations. For example, the phase transition induced by an added $X$-field can be detected by $\langle m_X\rangle$, as seen in Fig.~\ref{fig:2d}. However, Fig.~\ref{fig:2d} also shows that neither $\langle m_X\rangle$ nor $\langle m_Y\rangle$ can resolve the entire phase diagram alone. This is to be expected, as SPT phases cannot be completely detected by local order parameters. The SPEE, on the other hand, is inherently non-local, and can resolve the entire phase diagram. Nevertheless, the SPEE is still comparably simple to calculate given a PEPS ground state, so it is a genuinely useful tool to detect SSPT phase transitions.

Finally, we note a distinct property that the SPEE has as a tool to detect SSPT order. Usual quantities used to detect SPT order, such as string order parameters, are explicitly defined in terms of a certain symmetry \cite{Bartlett2010,Haegeman2012,Pollmann2012a,Marvian2017,You2018}. Such quantities therefore can only determine whether a state has SPT order with respect to this symmetry.  In contrast, the SPEE is ``symmetry-agnostic'', meaning that it is not defined with respect to any particular symmetry of the state \footnote{The geometry of the bipartition we use to calculate entropy is informed by the geometry of the subsystem symmetries, but this is the only way in which the symmetry enters the definition of the SPEE}. This is the reason that we were able to detect the phase of matter in Fig.~\ref{fig:2d} without a prior understanding of the relevant symmetry. The fact that a symmetry-agnostic quantity like the SPEE can characterize SSPT order reflects its fundamental differences from standard SPT order \cite{Schmitz2019}.

\section{3D cluster states} \label{sec:3d}

We have now seen that the 2D cluster phase can be detected and characterized by a value $\gamma=1$ of the SPEE. In this section, we move to investigating the SPEE in 3D systems. Going to 3D allows us to consider different types of subsystem symmetry, and also brings our discussion closer to fracton order, which exists only in dimension 3 and higher. We will consider cluster states defined on different 3D lattices, each with a different notion of subsystem symmetry, and each leading to a different type of area law correction. In light of our results for the 2D cluster phase, we expect the observed behaviour to hold throughout the corresponding SSPT phases.

\begin{figure}
    \centering
    \includegraphics[width=0.9\linewidth]{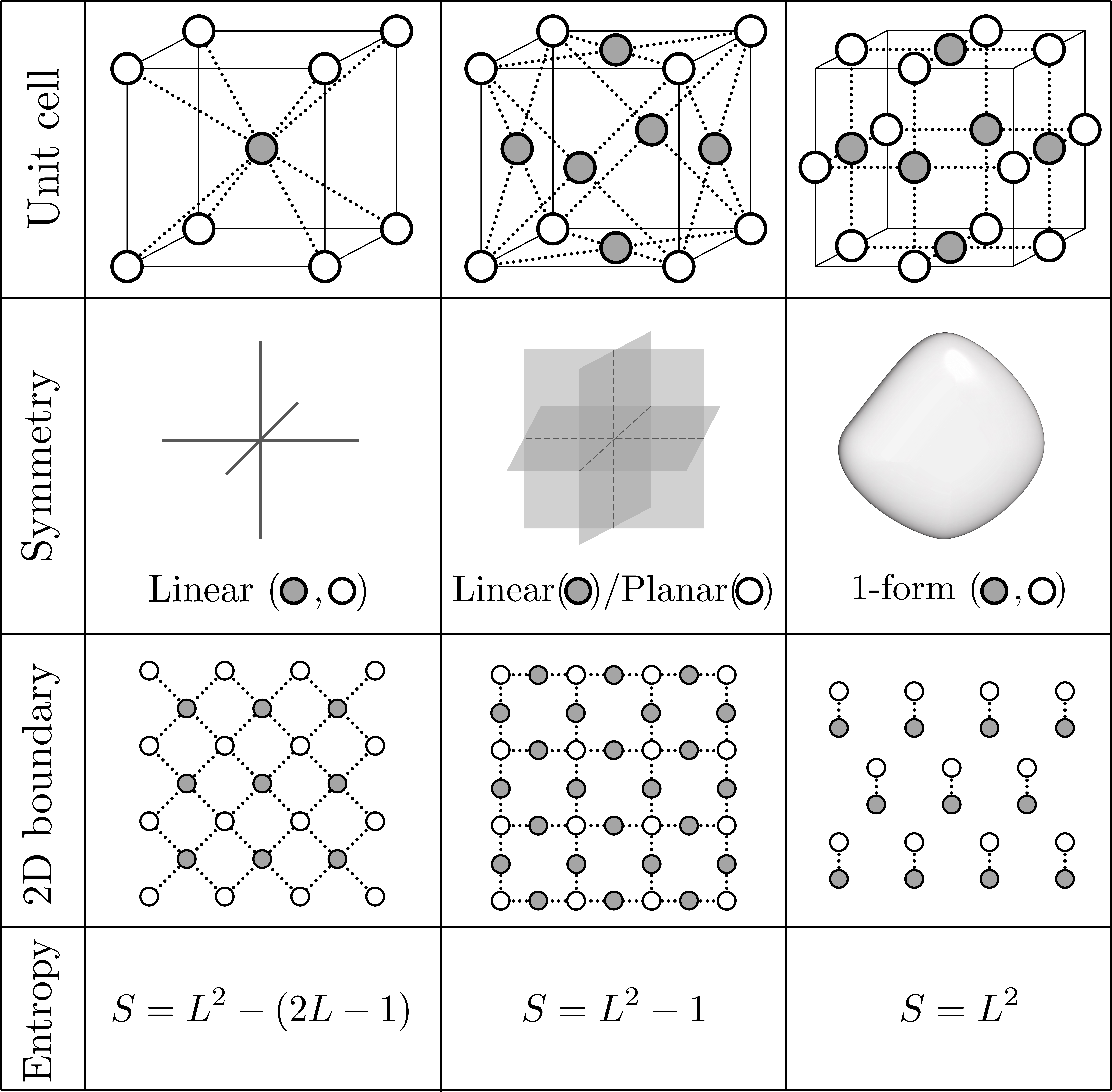}
    \caption{Summary of the results of Sec.~\ref{sec:3d}. The first row shows the unit cell of each 3D lattice, where light (dark) qubits lie on the A (B) sublattices, and graph state edges lie along dotted lines only. The second row illustrates the nature of the symmetries acting on each sublattice. The third row shows the effective 2D system living on he boundary of a planar bipartition, where light (dark) qubits live in the A (B) subsystems. The final row shows the entropy across this bipartition, where the boundary has dimensions $L\times L$.}
    \label{fig:3d}
\end{figure}

We start by reviewing how to define cluster states on lattices other than the 2D square lattice, also known as graph states. Consider a graph $\mcal{G}=(V,E)$ consisting of vertices $v\in V$ and edges $e\in E$. To define the corresponding graph state $|\mcal{G}\rangle$, we place a qubit in the $|+\rangle$ state on every vertex, and then act with $CZ$ on every pair of vertices connected by an edge,
\begin{equation}
    |\mcal{G}\rangle=\prod_{e\in E} CZ_e \bigotimes_{v\in V} |+\rangle_v.
\end{equation}
$|\mcal{G}\rangle$ is the ground state the following Hamiltonian:
\begin{equation}
    \mathcal{H}_\mcal{G}=-\sum_{v\in V} X_v \prod_{u\in N(v)} Z_u
\end{equation}
where $N(v)\subset V$ is the neighbourhood of $v$, and contains all vertices $u$ where are connected to $v$ by an edge. 

To compute entanglement entropy, we need only consider an effective 2D system lying along the boundary of the partition, as discussed in Sec.~\ref{sec:cluster}.
For these 2D boundary systems, we can use Eq.~(\ref{eq:stabentropy}) to calculate $S_A$. In what follows, we will consider 3D cubic lattices with dimensions $L\times L\times L$ (such that $|\partial A|=L^2$) and periodic boundary conditions in all directions. Similar statements can be made for open boundary conditions after the addition of boundary terms in the Hamiltonians. We will consider a biparition whose boundary is perpendicular to a coordinate axis, and all symmetries discussed will be tensor products of Pauli-$X$ operators. We note that graph state symmetries made of Pauli-$Y$ operators can also exist, but with drastically different, \textit{e.g.} fractal, geometry as shown in Ref.~\cite{Stephen2019}.

Our results are summarized in Fig.~\ref{fig:3d}. We will now briefly discuss each model.

\vspace{5mm}
\noindent
(\textit{i}) \textit{bcc lattice.} This model has subsystem symmetries in the form of lines moving in the $x,y,z$ directions of the lattice. It was previously studied in Ref.~\cite{You2018}, where it was also shown that it has non-trivial SSPT order under these linelike symmetries. The 2D system living on the boundary of the bipartition is exactly the 2D cluster state, where the $A$ and $B$ subsystems correspond to the $A$ and $B$ sublattices. Therefore, the $2L-1$ linelike symmetries of the cluster state generate $G_A$ in Eq.~(\ref{eq:stabentropy}), and we get a lower dimensional ``perimeter law'' correction to the area law.

\vspace{5mm}
\noindent
(\textit{ii}) \textit{fcc lattice.} The second model has linelike symmetries on the face qubits, but only planar symmetries on the vertex qubits. The 2D boundary system has only a single global symmetry in $G_A$, so there is a constant correction rather than a perimeter law correction. Thus, the planar symmetries seem to dictate the physics of this model, rather than the linelike symmetries. This example, along with the previous, suggest that SSPT phases in dimension $D$ with $k$-dimensional subsystem symmetries are associated with a correction to the area law that scales like $L^{D-k-1}$. 

\vspace{5mm}
\noindent
(\textit{iii}) \textit{RBH lattice.} The third model, first introduced by Raussendorf, Bravyi, and Harrington (RBH) \cite{Raussendorf2005}, was originally used in the context of fault-tolerant quantum computation \cite{Raussendorf2006,Raussendorf2007}. It was later shown in Ref.~\cite{Roberts2017} to have SPT order protected by so-called ``1-form'' symmetries \cite{Yoshida2016}. Such symmetries are distinct from subsystem symmetries because they are \textit{deformable}. While subsystem symmetries have rigid geometry like lines or planes, 1-form symmetries in a 3D system can live on \textit{any} closed 2D surface. This follows from local symmetries generated by applying $X$ to every face qubit in a single unit cell (or every edge qubit connected to a given vertex). In particular, this model also has planar symmetries as in the fcc lattice. Despite this, the boundary state is trivial, consisting of entangled pairs between the two subsystems, so there is no correction to the area law. The reason for this is as follows: For the previous two cases, as well as the 2D cluster state, the correction emerges because there is a non-local stabilizer that lives in one subsystem and is composed of local stabilizers that have support in both subsystems. For the case of 1-form symmetries, this is not the case: the non-local stabilizers can be decomposed into local stabilizers that are also contained in one subsystem. This behaviour was also noticed in a 2D example (which is actually the same state emerging on the boundary of the fcc lattice), in Ref.~\cite{Williamson2019}. Therefore, 1-form symmetries are likely not associated with area law corrections in general.

\section{Outlook} \label{sec:con}

In this paper, we have shown that the 2D cluster phase can be characterized by a uniform correction $\gamma=1$ to the area law which we have called the symmetry-protected entanglement entropy (SPEE), and we used this to construct a new numerical technique to detect SSPT order. The SPEE is relatively easy to calculate given a ground state PEPS tensor, hence it is an effective tool to detect SSPT order and SSPT phase transitions. A next step would be to check if the results presented in this paper are unique to the 2D cluster phase, or are generally true for all non-trivial SSPT phases. For example, Ref.~\cite{Devakul2018b} showed that the SSPT phases protected by the linelike symmetries considered here can be sorted into 8 different equivalence classes, and constructed representative states for each class. It is also easy to generalize the cluster state to different symmetry groups as in Ref.~\cite{You2018} and Ref.~\cite{Brell2015}, for example. Also, as shown here, there is rich behaviour of the correction for 3D cluster states. In all of these cases, it would be interesting to see whether there is a uniform value of the SPEE within the corresponding SSPT phases.

Using our method, we uncovered the surprising result that the SSPT order of the cluster state is preserved under the addition of local magnetic fields pointing anywhere in the $X$-$Y$ plane, as indicated by the SPEE. This indicates that the cluster phase defined previously \cite{Raussendorf2019,You2018,Devakul2018b} is part of a larger, more robust SSPT phase of matter. We attributed this larger phase of matter to subsystem time-reversal symmetry, although an in-depth understanding is still missing. In particular, the physical meaning of our notion of subsystem time-reversal is unclear. Interestingly, the newly discovered phase (`T') contains at least part of both the cluster phase (`X') and the fractal SSPT phase (`Y') containing the cluster state \cite{Stephen2019}. This suggests that the phases `X' and `Y', whose symmetries differ drastically in geometry, can potentially be unified by considering the anti-unitary symmetries of the phase `T'. 

This newly discovered phase may also have implications in terms of quantum computation. It is known that phases `X' and `Y' are computationally universal \cite{Raussendorf2019,Stephen2019}, meaning that every state within these phases may be used as a resource for universal measurement-based quantum computation. If the phase `T' is also computationally universal, this would mean that the computational power of the cluster state is robust to a much larger class of perturbations than previously thought. In general, it would be worthwhile to understand whether there is a clear link between the SPEE and the usefulness of a state as a computational resource, and if so, how well the SPEE compares to other figures of merit used to determine the computational usefulness of, \textit{e.g.}, perturbed cluster states \cite{Doherty2009,Kalis2012,Orus2013}.

In 3D, certain models with subsystem symmetries are dual to models with fracton order \cite{Vijay2016,Williamson2016,Kubica2018,You2018a,Shirley2019,Song2019}. This duality is realized by gauging the subsystem symmetries. In particular, if this gauging procedure is applied to a state with SSPT order, the resulting fracton model can be twisted \cite{You2018a}, in the same way that gauging a 2D SPT leads to a twisted Toric Code, \textit{i.e.} Double Semion model \cite{Levin2012}. It is therefore plausible to think that a transition between different fracton orders could be dual to a transition between different SSPT orders, which could in turn be detected by the SPEE, or a 3D generalization thereof. 

Finally, we briefly comment on the possibility to measure the SPEE experimentally. Measuring the topological entanglement entropy would be a direct way of observing the presence of topological order in a many-body state. However, this is experimentally daunting; a principle barrier being the difficulty of creating topological states in the first place. On the other hand, cluster states are relatively easy to make in optical lattices with nearest-neighbour interactions \cite{Mandel2003}. Furthermore, the second R\'enyi entropy can be measured in optical lattices by interfering two identical copies of a ground state and performing local parity measurements \cite{Abanin2012,Daley2012}, and this has been done to observe area laws in 1D systems \cite{Islam2015}. Thus, measuring the SPEE of the cluster phase seems feasible with current or near-term technologies. This would serve as a first route to verifying non-trivial quantum order via entanglement entropy.

\acknowledgments
DTS was supported by a fellowship from the Natural Sciences and Engineering Research Council of Canada (NSERC). 
This work has received funding from the European Research
Council (ERC) under the European Union’s Horizon 2020 research and innovation programme through the ERC Starting Grant WASCOSYS (No.~636201), and 
by the Deutsche Forschungsgemeinschaft (DFG) under Germany’s Excellence Strategy
(EXC-2111 -- 390814868).

\bibliographystyle{apsrev4-1}
\bibliography{biblio}

\onecolumngrid
\appendix

\section{Subsystem symmetry fractionalization in PEPS} \label{app:sspt}

In this appendix, we show how the non-trivial SSPT order of the cluster phase may be seen analytically by observing symmetry fractionalization on the PEPS boundary. Symmetry fractionalization, referring to the notion that the symmetry of an SPT ordered state acts in a fundamentally different way on the boundary of the system compared to in the bulk, is the basic principle of SPT order. For example, the boundary of 1D SPTs transforms protectively under the symmetry group \cite{Pollman2010,Chen2011,Schuch2011}, while the symmetry of 2D SPTs acts non-locally on the boundary \cite{Chen2011a}. Understanding how the symmetry acts on the boundary is the key to characterizing and classifying SPT order \cite{Chen2013,Else2014}. The case is the same for SSPT order, as discussed in Refs.~\cite{You2018,Devakul2018b}. For example, it was observed that, in the cluster phase, two parallel lines of subsystem symmetry which commute in the bulk may not commute on the boundary. This observation led to a classification of SSPT phases \cite{Devakul2018b}.

The framework of PEPS gives a natural way to understand the boundary of a many-body state.  Specifically, for a PEPS defined on a region with boundaries, the 
uncontracted virtual legs at the boundary of the region 
allow to parametrize the Hilbert space for
both the physical boundary excitations~\cite{Yang2014} and the entanglement spectrum of a bipartition~\cite{Cirac2011} on equal footing, and allow for a direct understanding of the way in which physical symmetries act on the boundary and entanglement.

\begin{figure}
    \centering
    \includegraphics[width=0.7\linewidth]{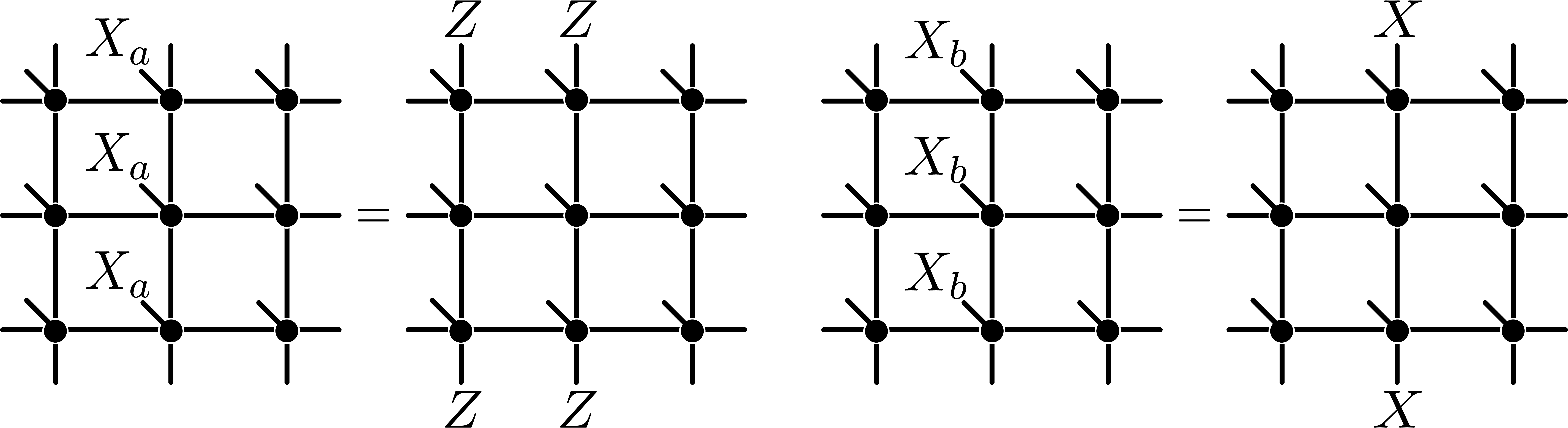}
    \caption{Action of the subsystem symmetries on the edge of a PEPS in the cluster phase. Therein, $X_a$ and $X_b$ denote the Pauli-$X$ operator acting on the $a$ and $b$ qubit of each lattice site as labelled in Fig,~\ref{fig:tn}.}
    \label{fig:edgesymms}
\end{figure}

In Fig.~\ref{fig:edgesymms}, we show how the subsystem symmetries act on these boundary degrees of freedom throughout the entire cluster phase. The relations pictured therein can be straightforwardly verified for the cluster state by using the definition of the tensor $C_2$. Looking just at the top edge of Fig.~\ref{fig:edgesymms}, we see that the boundary action of a given symmetry line anti-commutes with those of the neighbouring lines to the left and right (note that the $b$ qubit on each site is to the right of the $a$ qubit, cf. Fig.~\ref{fig:tn}), and commutes with all others. This projective representation has irreducible dimension $2^{N-1}$.
This same projective symmetry representation appears on the boundary of every state in the cluster phase, where the Pauli operators in the virtual space are acting on the protected subspace and the action on the junk subspace is trivial. One way to see this is to use Eq.~(\ref{eq:tensordecomp2}). A more elementary argument starts by adopting a quasi-1D picture of our 2D system by wrapping it onto a long cylinder. Then, the fact that the projective representation on the boundary cannot change without crossing a phase transition follows from the standard arguments for 1D SPT phases \cite{Pollman2010,Chen2011,Schuch2011}.

This symmetry fractionalization is a direct indication of the non-trivial SSPT order of the cluster phase. It can be used to derive, for example, the extensive ground state degeneracy of SSPT ordered systems with open boundaries \cite{You2018}, as well as a $2^{N-1}$-fold degeneracy in the entanglement spectrum on a cylinder of circumference $N$ \cite{Pollman2010}. The latter implies a bound on the entropy, $S_A\geq N-1$. Furthermore, our proof in Sec.~\ref{sec:anal} is based on these symmetries, which appear as symmetries of the transfer matrix in Eqs.~(\ref{eq:tmsymm1})-(\ref{eq:tmsymm2}). Therefore, the value $\gamma=1$ of the SPEE is a result of the way in which the subsystem symmetries fractionalize on the boundary.

Fig.~\ref{fig:edgesymms} is the PEPS equivalent of the picture developed in Ref.~\cite{Devakul2018b}, where the action of subsystem symmetries on a physical boundary are considered, as opposed to the virtual boundary considered here. The advantage of PEPS is that we can clearly see the same anti-commutation for all states in the cluster phase, and we are not restricted to models with zero correlation length. Different SSPT phases under the same subsystem symmetries would correspond to different anti-commutation relations on the virtual boundary. It should therefore be possible to reproduce the complete classification of Ref.~\cite{Devakul2018b} in the PEPS picture.

\section{Perturbation theory in PEPS} \label{app:perturbation}

In this appendix, we use perturbation theory in PEPS to show that states in the non-trivial phase of Fig.~\ref{fig:2d} may be well approximated by tensors of the form Eq.~(\ref{eq:perturbedA}). Consider a small perturbation away from the cluster state
\begin{equation} \label{eq_appendix_hamiltonian}
     \mathcal{H}=\mathcal{H}_C+\sum_{x,y} h'_{x,y},
\end{equation}
where $h'_{x,y}=-h_X X_{x,y}-h_Y Y_{x,y}$ with $h_X, h_Y \ll 1$. Throughout this appendix, we will label sites by a single index $j=(x,y)$ to condense notation. Defining
\begin{equation} \label{eq_appendix_V}
V (h_X,h_Y)=(\mathcal{H}_C-E_C)^{-1} (\mathbb{1} - \ket{C}\bra{C}) \sum_{j} h'_{j},
\end{equation}
the ground state, to first order in perturbation, is given by
\begin{align}
\label{eq_appendix_computation}
 \ket{\psi(h_X,h_Y)}&=\left[\mathbb{1} + V (h_X,h_Y)\right]\ket{C} \\ &= \left[\mathbb{1}-\sum_{j}(\mathcal{H}_C-E_C)^{-1} (\mathbb{1} - \ket{C}\bra{C}) \left( h_X X_{j} + h_Y Y_{j} \right)\right] \ket{C} \label{eq:app1}\\
&= \left[\mathbb{1}-\sum_{j}(\mathcal{H}_C-E_C)^{-1} \left( h_X X_{j} + h_Y Y_{j} \right) \right] \ket{C} \label{eq:app2}\\
&= \left[\mathbb{1} -\sum_{j} \left( \frac{h_X}{8} X_{j} + \frac{h_Y}{10} Y_{j} \right) \right]\ket{C} \label{eq:app3}\\
&= \Lambda (h_X,h_Y)^{\otimes N} \ket{C} + \mathcal{O}\left(h_X^2, h_Y^2, h_X h_Y \right) \label{eq:app4}
\end{align}
for 
\begin{equation} \label{eq_appendix_lambda}
\Lambda (h_X,h_Y)=\mathbb{1} - \frac{h_X}{8} X - \frac{h_Y}{10} Y.
\end{equation}
In going from Eq.~\eqref{eq:app1} to \eqref{eq:app2}, we have used that (adopting the notation of Sec.~\ref{sec:3d}),

\begin{align} \label{eq_appendix_orthogonal_1}
    \braket{C | X_j | C} &= \bra{+}^{\otimes N} \left(\prod_{e} CZ_e\right) X_{j}
    \left(\prod_{e'} CZ_{e'}\right) \ket{+}^{\otimes N} \\
    &= \bra{+}^{\otimes N} \left(\prod_{e} CZ_e\right)  \left(\prod_{e'} CZ_{e'}\right)
    X_{j} \prod_{j'\in N(j)} Z_{j'}
    \ket{+}^{\otimes N} \\
    &=\braket{+|+}^{N-4} \braket{+ |-}^4 \\
    &=0
\end{align}

and similarly $\braket{C | Y_{j} | C} = 0 \quad \forall j$, since 

\begin{equation} \label{eq_appendix_orthogonal_2}
Y_{j} \left(\prod_{e} CZ_e\right)
     = \left(\prod_{e} CZ_e\right) Y_{j} \prod_{j'\in N(j)} Z_{j'}
\end{equation}
To go from Eq.~\eqref{eq:app2} to \eqref{eq:app3}, we exploited the fact that
\begin{align} \label{eq_appendix_orthogonal_1}
    -H_C X_{j} \ket{C} &= \sum_{i} K_{i} X_{j} \ket{C} \\
    &=\sum_{i\in N(j)} K_{i} X_{j} \ket{C} 
    +\sum_{i\notin N(j)} K_{i} X_{j} \ket{C} \\
    &=-\sum_{i\in N(j)} X_{j} K_{i}  \ket{C} 
    +\sum_{i\notin N(j)} X_{j} K_{i}  \ket{C} \\
    &=4\sum_{i\in N(j)} X_{j}  \ket{C} 
    +(4-N)\sum_{i\notin N(j)} X_{j}  \ket{C}
\end{align}
which implies that for all $j$,  $X_{j}  \ket{C}$ is an eigenstate of $(\mathcal{H}_C-E_C)^{-1}$ with inverse energy 1/8 (note that $j \notin N(j))$. The same statement holds for $Y_{j}  \ket{C}$ with inverse energy 1/10.
The first order perturbation is implemented by acting on the cluster state by acting with $\Lambda$ on every site, cf. Eq.~(\ref{eq:clambda}). Note that $\Lambda$ fulfills $X \Lambda X = \Lambda^*$, by virtue of only consisting of Pauli-$X$ and $Y$ matrices with real coefficients. 

To reach higher orders in perturbation theory, we use the exponential perturbation theory developed in Ref.~\cite{Vanderstraeten2017}, which requires us to evaluate $\exp{(V)}|C\rangle$. Therefore, we must evaluate terms like $V^2 \ket{C}, V^3 \ket{C}$, and so on. We can use the machinery of \cite{Vanderstraeten2017} to write down PEPOs enacting the perturbations with increasing bond dimension for any given order. The extra bond dimension is needed to account for the fact that higher powers of $V$ will introduce products of $X$s and $Y$s with different prefactors, depending on the geometry of the operator. For example, the second order contribution to the ground state is given by
\begin{align}
\label{eq_appendix_second_order_ground_state}
& \left[ \frac{h_X^2}{16} \left( \frac{1}{8} \sum_{|i-j|=1} X_i X_j + \frac{1}{4} \sum_{|i-j|=\sqrt{2}} X_i X_j + \frac{1}{6} \sum_{|i-j|=2} X_i X_j + \frac{1}{8} \sum_{|i-j|>2} X_i X_j \right) \right. \nonumber \\
&\left. +\frac{h_Y^2}{20} \left( \frac{1}{6} \sum_{|i-j|=1} Y_i Y_j + \frac{1}{6} \sum_{|i-j|=\sqrt{2}} Y_i Y_j + \frac{1}{8} \sum_{|i-j|=2} Y_i Y_j + \frac{1}{10} \sum_{|i-j|>2} Y_i Y_j \right) \right. \\
&\left.  +\frac{9 h_X h_Y}{80} \left( \frac{1}{7} \sum_{|i-j|=1} X_i Y_j + \frac{1}{4} \sum_{|i-j|=\sqrt{2}} X_i Y_j + \frac{1}{7} \sum_{|i-j|=2} X_i Y_j + \frac{1}{9} \sum_{|i-j|>2} X_i Y_j \right) \right. \nonumber\\
&\left.  -\frac{h_X h_Y}{80} \sum_i X_i Y_i \right] \ket{C} \nonumber,
\end{align}
where the first four terms sum over nearest neighbours, next-nearest neighbours, next-next-nearest neighbours, and all other pairings, respectively. Together with the first order correction, one (not necessarily optimal) PEPO  that implements this correction is given by a local tensor with bond dimension $D=11$ and non-zero entries (the order of the virtual legs is north, west, south, east) given by
\begin{align}
\label{eq_appendix_second_order_tensor}
B_{0,0,0,0} &= \mathbb{1} + a X + b Y + c XY \nonumber \\
B_{0,0,0,1} &= B_{0,0,1,0} = B_{0,2,0,0} = B_{2,0,0,0} = d X \nonumber \\
B_{0,1,0,2} &= B_{1,0,2,0} = \mathbb{1} \nonumber \\
B_{1,1,0,0} &= B_{0,0,2,2} = B_{1,0,0,2} = B_{0,1,2,0} = e \mathbb{1} \nonumber \\
B_{0,0,0,3} &= B_{0,0,3,0} = B_{0,3,0,0} = B_{3,0,0,0} = f Y \nonumber \\
B_{0,4,0,0} &= B_{0,0,0,5} = B_{4,0,0,0} = B_{0,0,5,0} = g Y \nonumber \\
B_{0,5,0,4} &= B_{5,0,4,0} = \mathbb{1}  \\
B_{4,4,0,0} &= B_{4,0,0,5} = B_{0,0,5,5} = B_{0,4,5,0} = h \mathbb{1} \nonumber \\
B_{0,0,0,6} &= B_{0,0,6,0} = B_{0,7,0,0} = B_{7,0,0,0} = k_X X  \nonumber \\
B_{0,0,0,7} &= B_{0,0,7,0} = B_{0,6,0,0} = B_{6,0,0,0} = k_Y Y \nonumber \\
B_{0,0,0,8} &= B_{0,8,0,0} = B_{11,0,0,0} = B_{0,11,0,0} = l_X X  \nonumber \\
B_{0,9,0,0} &= B_{9,0,0,0} = B_{0,0,10,0} = B_{0,0,0,10} = l_Y Y \nonumber \\
B_{0,8,0,9} &= B_{8,0,9,0} = B_{0,10,0,11} = B_{10,0,11,0} =\mathbb{1} \nonumber \\
B_{10,8,0,0} &= B_{8,10,0,0} = B_{0,8,9,0} = B_{0,10,11,0} = B_{8,0,0,9} \nonumber \\ &= B_{9,0,0,11} = B_{0,0,11,9} = B_{0,0,10,11} = m \mathbb{1} \nonumber 
\end{align}
\begin{align*}
a &= d = \frac{h_X}{8}, & b &= \frac{h_Y}{10}, & c &= -\frac{h_X h_Y}{80}, & e  &=- \frac{191}{192}, & f &= \frac{h_Y}{16} \sqrt{\frac{2}{3}}, \nonumber \\
g &= \frac{h_Y}{20}, & h &= \frac{8}{3}, & k_X &= l_X = \frac{h_X}{280}, & k_Y &= l_Y = h_Y, & m&=\frac{35}{8}, \nonumber
\end{align*}
which fullfils $X B_{ijkl} X = B^*_{ijkl}$ for all $i,j,k,l$. The tensor $B$ therefore commutes locally with $X$, up to complex conjugation, as in Eq.~(\ref{eq:perturbedA}).

We will now argue why this holds to all orders in perturbation theory: As long as the action of $V^n \ket{C}$ on the cluster state can be expressed as sums of Pauli strings of $X$s and $Y$s with real coefficients, one can always find a PEPO $B$ for $\exp(V)$ which fulfills $X B_{ijkl} X = B^*_{ijkl}$. In other words, we want to show that
\begin{align}
    \label{eq_appendix_hypothesis}
    V^n\ket{C} = \sum_{\mbf{a}  \mbf{b}} c_{\mbf{a} \mbf{b}} P_{\mbf{a}  \mbf{b}}  \ket{C}
\end{align}
for all $n$, where $P_{\mbf{a} \mbf{b}} = \bigotimes_{k} X^{a_k} Y^{b_k}$, $\mbf{a}$ and $\mbf{b}$ are bitstrings on the lattice and the $c_{\mbf{a} \mbf{b}}$ are real coefficients. We have seen explicitly that Eq.~(\ref{eq_appendix_hypothesis}) holds for $n=1,2$. Now, assuming  Eq.~(\ref{eq_appendix_hypothesis}) holds for some $n$, we have,
\begin{align}
    \label{eq_appendix_proof}
    V^{n+1}\ket{C}&=V \sum_{\mbf{a}  \mbf{b}} c_{\mbf{a} \mbf{b}} P_{\mbf{a}  \mbf{b}}   \ket{C} = (\mathcal{H}_C-E_C)^{-1} (\mathbb{1} - \ket{C}\bra{C}) \sum_{x,y} h'_{x,y} \sum_{\mbf{a}  \mbf{b}} c_{\mbf{a} \mbf{b}} P_{\mbf{a}  \mbf{b}} \ket{C} \nonumber\\
    &=  \sum_{\mbf{a}  \mbf{b}} c'_{\mbf{a} \mbf{b}} P_{\mbf{a}  \mbf{b}} \ket{C},
\end{align}
where $c'_{\mbf{a} \mbf{b}}$ are also real. Therein, we have used that the perturbation $h'$ consists of Pauli-$X$s and $Y$s with real coefficients and the fact that each product of Pauli matrices either maps the cluster state to an exact excitation or to itself, making $(\mathcal{H}_C-E_C)^{-1} (\mathbb{1} - \ket{C}\bra{C})$ act simply as a multiplication of each of the $c_{\mbf{a} \mbf{b}}$ by a real number. So Eq.~(\ref{eq_appendix_hypothesis}) holds for all $n$ by induction.

Therefore, it is clear that, at any given order, the corresponding PEPO-tensor $B$ describing the perturbation will again act as a real linear combination of $\mathbb{1}, X$, $Y$ and $XY$ for any given virtual state, hence the resulting PEPS tensor has the form of Eq.~(\ref{eq:perturbedA}). We conclude that there must exist a region in $(h_X, h_Y)$ space around the cluster point in which the ground state is accurately described by a PEPO fulfilling Eq.~(\ref{eq:perturbedA}) acting on the cluster state.

\end{document}